\documentclass[12pt]{article}

\usepackage{hyperref}

\makeatletter \@addtoreset{equation}{section} \makeatother

\def\be{\begin{equation}}
\def\ee{\end{equation}}
\def\ba{\begin{array}}
\def\ea{\end{array}}
\def\d{\partial}
\def\ga{\alpha}
\def\dps{\displaystyle}

\newcommand{\half}{\frac{1}{2}}

\begin{document}

\begin{flushright}
{\tt FIAN/TD/11/03}
\end{flushright}

\vspace{1cm}

\begin{center}

{\bf \Large On the Frame-Like Formulation of
Mixed-Symmetry Massless  Fields in $(A)dS_d$}

\vspace{.7cm}

\textsc{K.B. Alkalaev, O.V. Shaynkman and M.A. Vasiliev}

\vspace{.7cm}

{\em I.E.Tamm Department of Theoretical Physics, P.N.Lebedev Physical
Institute,\\
Leninsky prospect 53, 119991, Moscow, Russia}

\vspace{3mm}
E-mails: ${\tt alkalaev@lpi.ru \quad  shayn@lpi.ru \quad
vasiliev@lpi.ru}$

\vspace{3mm}

\vspace{1cm}

\begin{abstract}

The frame-like covariant Lagrangian formulation of bosonic and
fermionic mixed-symmetry type higher spin massless fields propagating
on the $AdS_d$ background is proposed. Higher spin fields
are described in terms of gauge $p$-forms which carry
tangent indices representing certain traceless tensor
or gamma transversal spinor-tensor
representations of the $AdS_d$ algebra  $o(d-1,2)$
(or $o(d,1)$ for bosonic fields in $dS_d$).
Manifestly gauge invariant Abelian higher spin
field strengths are introduced for the general case. We
describe the general framework and demonstrate how it works
for the mixed-symmetry type fields associated with the
three-cell ``hook"  and arbitrary two-row rectangular
tableaux. The manifestly  gauge invariant actions for these
fields are presented in a simple form. The flat limit is
also analyzed.

\end{abstract}

\end{center}

\vspace{1cm}

\section{Introduction}

The problem of covariant Lagrangian description of arbitrary spin
fields propagating on flat \cite{PF}-\cite{hull} and
(anti)de Sitter ($(A)dS$)
\cite{Fronsdal_ads}-\cite{Zinoviev2} backgrounds attracts
considerable attention. The interest is motivated
by the fact that higher spin fields and higher spin symmetries
show up in a wide range of models from string theories to higher
spin gauge theories describing interacting dynamics of massless
fields on the Minkowski \cite{pos}-\cite{FM}
and the $(A)dS_d$ backgrounds
\cite{AdS_interact}-\cite{vas_int} (for review and more
references see \cite{V_obz2,vas_int}).
Possible relations of higher spin gauge theory with a
tensionless limit of string theory in $AdS$ space
and boundary conformal models was extensively discussed in
\cite{Sundborg:2000wp}-\cite{Schnitzer:2003zr}
in the context of weak coupling regime of the $AdS/CFT$ correspondence
\cite{Maldacena:1997re}-\cite{Witten:1998qj}.

To date, symmetric higher spin field dynamics (both massive
and massless) provides the most elaborated case among the
variety of unitary irreps of Poincare and $AdS$ algebras
\cite{SH,Fronsdal,vas_yadfiz,Fronsdal_ads,V1,LV,vf,Deser}.
To some extent, this is because in the four dimensional
space-time there is no room for mixed-sym\-metry irreps
except for dual theories involving ``exotic" symmetry type
dynamical variables\footnote{ For detailed discussion of
dual theories in diverse  dimensions see recent papers
\cite{hull2}, \cite{bekaert}, \cite{bekaert2},
\cite{hull1}, \cite{Casini}, \cite{henneaux}.}. However,
for higher space-time dimensions, mixed-symmetry
representations do appear and the problem of their
field-theoretical description has not been yet worked out
in full generality. In case of Minkowski space several
approaches were suggested to analyze mixed-symmetry fields
\cite{Labastida,Pashnev,bekaert,hull}. Covariant
formulation for generic mixed-symmetry fields in $AdS_d$ is
still lacking however, despite some progress achieved in
\cite{BMV,Metsaev_d5,siegel,Zinoviev,Zinoviev2}. The
peculiarity which complicates the straightforward extension
of the flat results is that the classification of massless
fields is essentially different for Poincare and $AdS_d$
algebras. {}From the field-theoretical perspective, this
fact manifests itself in different sets of gauge symmetries
in flat and $AdS_d$ space-times \cite{Metsaev,BMV}. As a
consequence, an irreducible $AdS_d$ mixed-symmetry field
decomposes into a set of flat fields in the flat limit
\cite{BMV}.

In this paper we propose a new approach to the covariant
description of generic mixed-symmetry  fields propagating on the
$AdS_d$ background, which generalizes the ``gauge" formulation of
the symmetric field dynamics developed previously in
\cite{vas_yadfiz,V1,LV,vf,d5} as well as analogous first-order
approach elaborated by Zinoviev in \cite{Zinoviev2} for particular
mixed-symmetry fields.

The construction is surprisingly simple.
Let  a lowest weight unitary massless representation of
the $AdS_d$ algebra $o(d-1,2)$ be characterized by the lowest
energy subspace
described as a representation of $o(d-1)\subset o(d-1,2)$
by a traceless Young tableau $Y_{o(d-1)}$ which has a longest
row of length $s$ and a shortest column of height $p$.
Then the corresponding field-theoretical system can be described
by a $p$-form gauge field which takes
values in the representation of the $AdS_d$ algebra $o(d-1,2)$
described by the traceless Young tableau $Y_{o(d-1,2)}$
obtained from that of $Y_{o(d-1)}$ by
cutting the shortest column and adding the longest
row of length $s-1$. The resulting $p$-form gauge field contains
the physical higher spin gauge field along with all necessary
auxiliary and extra fields and allows one to construct manifestly
gauge invariant field strengths to be used to build invariant
action in the MacDowell-Mansouri form \cite{MacDowell:1977jt}. The
formulation in terms of $p$-form connections and higher spin
curvatures allows us to control higher spin gauge symmetries in
geometric terms.

As far as bosonic massless fields are concerned, the proposed
formulation works equally well in de Sitter background. For
fermions this is not the case because the $dS$ reality
conditions for massless fields
require imaginary mass-like parameters in the action. For
definiteness we will mostly refer to the  $AdS$ case in this paper.

The paper is organized as follows. In section 2 we present the
general scheme, fix an appropriate set of fields and gauge
symmetries, discuss a form of the action functional and
generalized Weyl tensors. The particular examples of three-cell
``hook" tableau, four-cell ``window" tableau, and an arbitrary
two-row rectangular Young tableau
are considered, respectively, in subsections 3.1, 3.2 and  3.3 of
section 3. For these models we build manifestly gauge invariant
actions which properly describe the field dynamics on the
$(A)dS_d$ background and investigate their flat limits. Conclusion
is given in section 4.

\section{General scheme}

\subsection{Young tableaux and trace conditions}

Let $A^{(s_1, ... , s_q)}$ denote a tensor\footnote{Throughout
the paper we work within the mostly minus signature
and use notations $\underline{m},\underline{n} = 0\div d-1\;$ for
world indices,
$a,b= 0\div d-1$ for tangent Lorentz $so(d-1,1)$ vector indices and
$A,B = 0 \div d $ for tangent $(A)dS_d$ ($so(d-1,2)$)$so(d,1)$ vector
indices.
We also use condensed notations of \cite{V1} for a set of symmetric vector
indices:
$a (k)\equiv (a_1 \ldots a_k)$.
Upper (lower) indices denoted by the same letter are assumed to be
symmetrized as $X^a Y^a \equiv \frac{1}{2!}(X^{a_1}Y^{a_2}+
X^{a_2}Y^{a_1})$
prior contractions.}
\be
\label{field1}
A^{a_1(s_1),\,a_2(s_2),\, ...\,,\, a_{q}(s_{q})}\;,
\ee
which is  symmetric in each group of
indices\footnote{Usually, the parameter $q$ in (\ref{field1})
satisfies the inequality $q\leq \nu$, where $\nu= [\frac{d}{2}]-1$ is
the rank of the little group $SO(d-2)$ for Minkowski space
or $\nu=[\frac{d-1}{2}]$ is the rank of
$SO(d-1)$ in the case of $AdS_d$, although
dual descriptions with larger $q$ are also possible.}
$a_i (s_i )$ and satisfies the
Young symmetry conditions associated with the Young tableau
$Y(s_1 , s_2 , \ldots, s_q )$
composed of
$q$ rows of lengths $s_1 \geq s_2 \geq s_3\geq\ldots \geq s_q >0$,
{\it i.e.} symmetrization of all indices in $i^{th}$
row with any index from some $(i+k)^{th}$ $(k>0 )$ row
gives zero.

Let the tensor (\ref{field1}) satisfy  the conditions
\be
\label{tr1}
\eta_{a_i a_i} \eta_{a_i a_i}
A^{a_1(s_1),\,a_2(s_2),\, ...\,,\, a_{q}(s_{q})}=0\;, \qquad 0<  i \leq
m\,
\ee
and
\be
\label{tr2}
\eta_{a_i a_i}
A^{a_1(s_1),\,a_2(s_2),\, ...\,,\, a_{q}(s_{q})}=0\;, \qquad m<  i \leq
q\,,
\ee
where $\eta_{ab}$ is some metric, which has
signature $(p,r)$ (for the Lorentz case, for example, $p=d-1$
and $r=1$)
and $m$ is some
non-negative integer. The condition (\ref{tr1}) means that
contraction of any two pairs of indices from any of the first
$m$ rows of the Young tableau gives zero.
The condition (\ref{tr2}) means that contraction of any pair of
indices from the last $q - m$ rows gives zero.

The linear space of tensors (\ref{field1}) which
have the Young properties of the
type $Y(s_1, \ldots,  s_q )$ and satisfy
 the conditions (\ref{tr1}), (\ref{tr2}) will be denoted
$B^{p,r}_m(s_1, \ldots , s_q )$.
Note that $B^{p,r}_i(s_1, \ldots , s_q )
\subset B^{p,r}_j(s_1, \ldots , s_q )$ for $i<j$.
$B^{p,r}_0(s_1, \ldots , s_q )$ is the space of traceless tensors
with the $Y(s_1, \ldots, s_q )$ Young properties.

The following lemmas are simple consequences of the definitions
(\ref{tr1})-(\ref{tr2}) and the Young symmetry properties of
(\ref{field1}).

\vspace{3mm}

{\it  \underline{Lemma 1}}

Contraction of $\eta_{(a_ia_j}\eta_{a_ka_l)}$ with
any four symmetrized indices of a tensor from
$B^{p,r}_m(s_1, \ldots , s_q )$ gives zero.

Lemma 1 is a corollary of (\ref{tr1}) and the Young symmetry properties,
which guarantee that any group of symmetrized indices can be
placed in the first row.

{\it  \underline{Lemma 2}}

{}From Lemma 1 it follows that
\be
\label{tr3}
\eta_{a_i (a_j} \eta_{a_k a_l)}
A^{a_1(s_1),\,a_2(s_2),\, ...\,,\, a_{q}(s_{q})}=0\;, \qquad
\forall \;i,j,k,l\,,
\ee
\textit{i.e.} any double trace gives zero provided that any
three of the contracted indices  are symmetrized.

This is because $\eta_{ab} \eta_{cd}$
belongs to the symmetric part of the tensor product
\be
\Big(\begin{picture}(12,12)(-1,4)
{\linethickness{0.210mm}
\put(00,10){\line(1,0){10}} 
\put(00,05){\line(1,0){10}} 
\put(00,05){\line(0,1){5}} 
\put(05,05){\line(0,1){5}} 
\put(10,05){\line(0,1){5}}
}
\end{picture}
\otimes
\begin{picture}(12,12)(-1,4)
{\linethickness{0.210mm}
\put(00,10){\line(1,0){10}} 
\put(00,05){\line(1,0){10}} 
\put(00,05){\line(0,1){5}} 
\put(05,05){\line(0,1){5}} 
\put(10,05){\line(0,1){5}}
}
\end{picture}
\Big)_{\rm sym}
 =
\begin{picture}(22,12)(-1,4)
{\linethickness{0.210mm}
\put(00,10){\line(1,0){20}} 
\put(00,05){\line(1,0){20}} 
\put(00,05){\line(0,1){5}} 
\put(05,05){\line(0,1){5}} 
\put(10,05){\line(0,1){5}}
\put(15,05){\line(0,1){5}} 
\put(20,05){\line(0,1){5}} 
}
\end{picture}
\oplus
\begin{picture}(13,12)(-1,1)
{\linethickness{0.210mm}
\put(00,10){\line(1,0){10}}
\put(00,05){\line(1,0){10}}
\put(00,00){\line(1,0){10}}
\put(00,00){\line(0,1){10}}
\put(05,00){\line(0,1){10}}
\put(10,00){\line(0,1){10}}
}
\end{picture}\,.
\ee
Nonzero traces in $B^{p,r}_m(s_1, \ldots , s_q )$
therefore can only appear when all elementary contractions hit
different rows.

{\it  \underline{Lemma 3}}

The condition (\ref{tr2}) along with Lemma 2 mean that
contraction of any $m+1$ pairs of indices of
$A^{a_1(s_1),\,a_2(s_2),\, ...\,,\, a_{q}(s_{q})}\in
B^{p,r}_m(s_1, \ldots , s_q )$ gives zero.

\vspace{0.2cm}
It is convenient to treat Young tableaux as built
of horizontal
rectangular blocks $(s,p)$ of length $s$ and height $p$ as
elementary entities:
\be
\label{block}
\begin{picture}(130,55)(-1,40)
\put(50,96){$s$}
{\linethickness{.5mm}
\put(00,90){\line(1,0){100}}%
\put(00,50){\line(1,0){100}}%
\put(100,50){\line(0,1){40}}%
\put(00,50){\line(0,1){40}}%
}
\put(00,70){\line(1,0){100}}
\put(00,80){\line(1,0){100}}
\put(00,60){\line(1,0){100}}
\put(00,50){\line(1,0){100}}
\put(10,50.0){\line(0,1){40}}
\put(20,50.0){\line(0,1){40}}
\put(30,50.0){\line(0,1){40}}
\put(40,50.0){\line(0,1){40}}
\put(50,50.0){\line(0,1){40}}
\put(60,50.0){\line(0,1){40}}
\put(70,50.0){\line(0,1){40}}
\put(80,50.0){\line(0,1){40}}
\put(90,50.0){\line(0,1){40}}
\put(105,65){$p$}%
\end{picture}
\ee
Then one operates with a Young tableau  $Y(s_1, ... , s_q)$
with indices rearranged into elementary blocks
\be
\label{block_notation}
A^{(s_1\,,\, ... \, , s_q)} \sim A^{(\tilde{s}_1,p_1);(\tilde{s}_2,p_2);
\, ... \, ;\,(\tilde{s}_k,p_k)}\;,
\ee
where blocks are described by the sets of pairs of positive integers
$(\tilde{s}_i,p_i)$ with
$\tilde{s}_1 > \tilde{s}_2 > \cdots > \tilde{s}_k >0$ and $p_i$
such that $\sum_i p_i=q$. The exact identification in (\ref{block_notation}) is
\be
\label{block_notation2}
\tilde{s}_1= \underbrace{s_1= \ldots = s_{p_1}}_{p_1} >
\tilde{s}_2=\underbrace{s_{p_1+1} = \ldots = s_{p_1+p_2}}_{p_2}
> \ldots >\tilde{s}_k = \underbrace{s_{p_1+ \ldots +p_{k-1}+1} = \ldots =
s_q}_{p_k}\;.
\ee
For the upper block it is sometimes convenient to use notations
$\tilde{s}_1 =s$ and
$p_1=p$.

Recall that a rectangular block is invariant (may be up to a sign)
with respect to exchange of its rows.
As a result it follows

\vspace{0.2cm}
{\it  \underline{Lemma 4}}

Once  (\ref{tr2}) is true for one of the rows of a rectangular block it
is true for the entire block, i.e.
$B^{p,r}_{m_1}(s_1, \ldots , s_q )=B^{p,r}_{m_2}(s_1, \ldots , s_q )$
if $s_{m_1+1} = s_{m_2+1}$.

Therefore, it is sufficient to impose the trace condition (\ref{tr2})
for any row inside a horizontal block
({\it e.g.}, upper row).

\subsection{Background geometry and compensators}

The background Minkowski or $(A)dS_d$ geometry is described by
the frame field $h^a= h_{\underline{n}}{}^a\,dx^{\underline{n}}$ and
Lorentz spin connection
$\omega^{ab} = \omega_{\underline{n}}{}^{ab}\,dx^{\underline{n}}$
which obey the equation
\be
\label{backgrounds}
[{\cal D}_{\underline{m}},{\cal D}_{\underline{n}}]A^{a_1(s_1),\,
...\,,a_q(s_q)}
= \lambda^2 (s_1\,\,h_{\underline{m}}{}^{a_1}\, h_{\underline{n}\,c}
\,A^{a_1(s_1-1)c,\, ...\,,a_q(s_q)}+ \ldots)
- (\underline{m}\leftrightarrow\underline{n})\;,
\ee
where
\be
\label{Lorentzderiv}
{\cal D}_{\underline{n}}A^{a_1(s_1),\, ...\,,a_q(s_q)}
= \d_{\underline{n}}A^{a_1(s_1),\, ...\,,a_q(s_q)}
+s_1\,\omega_{\underline{n}}{}^{a_1}{}_c\,A^{a_1(s_1-1)c,\,
...\,,a_q(s_q)}+ \;\ldots  \;,
\quad \d_{\underline{n}}=\frac{\d}{\d x^{\underline{n}}}\;.
\ee
The zero-torsion condition
${\cal D}_{\underline{n}} h_{\underline{m}}{}^a
-{\cal D}_{\underline{m}} h_{\underline{n}}{}^a = 0$
is imposed. It expresses
$\omega_{\underline{m}}{}^{ab}$ in terms of $h_{\underline{m}}{}^a$.
Note that the equation (\ref{backgrounds}) describes $AdS_d$ space-time
with
the symmetry algebra $o(d-1,2)$ when $\lambda^2>0$ and $dS_d$
space-time with the
symmetry algebra $o(d,1)$ when $\lambda^2<0$. Minkowski space-time
corresponds to $\lambda=0$. In the fermionic case, massless equations
contain mass-like terms expressed in units of $\lambda$. A formal
complication for the de Sitter case is that these terms become
imaginary.

In the sequel, we extensively use $(A)dS_d$ covariant notations
and operate with Young tableaux
$A^{A_1(s_1),A_2(s_2), ... ,A_k(s_k)}(x)$, where $A_i = 0\div d$
is an $o(d-1,2)$ or $o(d,1)$ vector index. To relate the $(A)dS$
covariant approach with the Lorentz-covariant approach it is
useful to introduce the compensator field
$V^A(x)$ normalized as $V^A V_A = \pm\, 1$ \cite{compensator}. It allows
one to identify the Lorentz subalgebra $so(d-1,1)$ of the
$(A)dS_d$ algebra $(so(d-1,2))so(d,1)$ with the stability algebra of the
compensator. With the help of the compensator field, the covariant
splitting of the  $(so(d-1,2)) so(d,1)$ 1-form connection
$\Omega^{AB}=-\Omega^{BA}$ into the frame field $E^A$ and the Lorentz
spin connection $\omega^{AB}=-\omega^{BA}$ is defined as follows
\be
\label{defEOmega}
\lambda\,E^A = DV^A\equiv dV^A+\Omega^{AB}V_B\;,
\qquad \omega^{AB} = \Omega^{AB} \mp
\lambda\,(E^{A}\,V^{B}-E^{B}\,V^{A})\;.
\ee
It follows that
\be
E^A\,V_A = 0\;,
\quad
{\cal D}V^A = dV^A+\omega^{AB}V_B\equiv 0\;.
\ee
The metric tensor  is
$g_{\underline{mn}}=E_{\underline{m}}{}^A
E_{\underline{n}}{}^B\,\eta_{AB}\,.
$
In these notations, the background $(A)dS_d$
geometry  is described by the $(A)dS_d$ connection
$W^{AB}$=($h^a$, $\omega_0^{ab}$) satisfying the
zero-curvature equation (see, e.g., \cite{V_obz2} for more detail)
\be
\label{zerocurv}
R^{AB}(W)\equiv  dW^{AB}+W^A{}_C\wedge W{}^{CB}=0\;.
\ee
The action of the covariant derivative $D_0$ on an
arbitrary $(A)dS_d$ tensor is given by
\be
\ba{c}
\label{backderaction}
D_0 A^{A_1(s_1),\,  \ldots\, , A_k(s_k)} = d A^{A_1(s_1)\,,\, ... \,,\,
A_k(s_k)}
\\
\\
+ s_1 \,W^{A_1}{}_C\wedge A^{CA_1(s_1-1)\,,\, ... \,,\, A_k(s_k)} + \ldots
+ s_k \, W^{A_k}{}_C\wedge A^{A_1(s_1)\,,\, ... \,,\, CA_k(s_k-1)}\;.
\ea
\ee

\subsection{Mixed-symmetry bosonic massless fields}

Relativistic fields in $AdS_d$ which admit quantum-mechanically
consistent formulation
are classified according to  lowest weight unitary representations
of $o(d-1,2)$. Unitarity is the standard quantum mechanical
requirement while lowest weight guarantees that the energy is
bounded from below. Note that the case of $dS_d$ does not
allow irreps which are both unitary and lowest (highest) weight,
that makes important difference compared to $AdS_d$.

Lowest weight unitary irreps $D(E_0,{\bf s})$
are constructed in a standard
fashion starting with a vacuum space $|E_0,{\bf s}\rangle$
that forms a unitary module of
the maximal compact subalgebra
$o(2)\oplus o(d-1)\subset o(d-1,2)$. Here $E_0$ is lowest
energy eigenvalue  and ${\bf s}=(s_1, ... ,s_q,0, ..., 0)$ with
$q\leq \nu= [\frac{d-1}{2}]$
is a generalized spin. In terms of Young tableaux, $s_i$
is the length of $i^{th}$ row of the
$o(d-1)$ Young tableau $Y(s_1, ... ,s_q)$.

Let the vacuum representation of $o(d-1,2)$ with some energy $E_0$
form a finite-dimensional irrep of $o(d-1)$ characterized by
the $o(d-1)$ traceless Young tableau

\be
\label{vacuum}
\bigskip
\begin{picture}(185,350)%
\begin{picture}(109,300)%
{\linethickness{.500mm}
\put(30,32){\footnotesize $\tilde{s}_{k}$}
\put(50,10){\line(0,1){30}}
\put(00,10){\line(1,0){50}}
\put(00,10){\line(0,1){30}}
\put(00,40){\line(1,0){50}}}
\put(.0,40){\line(1,0){50}}
\put(.0,30){\line(1,0){50}}
\put(.0,20){\line(1,0){50}}
\put(10,10){\line(0,1){30}}
\put(20,10){\line(0,1){30}}
\put(30,10){\line(0,1){30}}
\put(40,10){\line(0,1){30}}
\put(55,20){\footnotesize $p_{k}$}
\end{picture}
\begin{picture}(125,90)(113,-30)%
{\linethickness{.500mm}
\put(30,62){\footnotesize $\tilde{s}_{k-1}$}
\put(80,10){\line(0,1){60}}
\put(00,10){\line(1,0){80}}
\put(00,10){\line(0,1){60}}
\put(00,70){\line(1,0){80}}}
\put(.0,60){\line(1,0){80}}
\put(.0,50){\line(1,0){80}}
\put(.0,40){\line(1,0){80}}
\put(.0,30){\line(1,0){80}}
\put(.0,20){\line(1,0){80}}
\put(10,10){\line(0,1){60}}
\put(20,10){\line(0,1){60}}
\put(30,10){\line(0,1){60}}
\put(40,10){\line(0,1){60}}
\put(50,10){\line(0,1){60}}
\put(60,10){\line(0,1){60}}
\put(70,10){\line(0,1){60}}
\put(85,30){\footnotesize $p_{k-1}$}
\end{picture}
\begin{picture}(75,75)(242,-90)%
{\linethickness{.5mm}
\put(00,10){\line(0,1){70}}}
\put(10,20){\circle*{2}}
\put(20,20){\circle*{2}}
\put(30,20){\circle*{2}}
\put(40,20){\circle*{2}}
\put(50,40){\circle*{2}}
\put(40,40){\circle*{2}}
\put(30,40){\circle*{2}}
\put(10,40){\circle*{2}}
\put(20,40){\circle*{2}}
\put(60,60){\circle*{2}}
\put(50,60){\circle*{2}}
\put(40,60){\circle*{2}}
\put(30,60){\circle*{2}}
\put(10,60){\circle*{2}}
\put(20,60){\circle*{2}}
\end{picture}
\begin{picture}(125,95)(321,-160)%
\put(30,81.5){\footnotesize $\tilde{s}_2$}%
{\linethickness{.500mm}
\put(120,10){\line(0,1){80}}
\put(00,90){\line(1,0){120}}
\put(00,10){\line(1,0){120}}
\put(00,10){\line(0,1){80}}}
\put(00,70){\line(1,0){120}}
\put(00,80){\line(1,0){120}}
\put(00,60){\line(1,0){120}}
\put(00,50){\line(1,0){120}}
\put(00,40){\line(1,0){120}}
\put(00,30){\line(1,0){120}}
\put(00,20){\line(1,0){120}}
\put(10,10.0){\line(0,1){80}}
\put(20,10.0){\line(0,1){80}}
\put(30,10.0){\line(0,1){80}}
\put(40,10.0){\line(0,1){80}}
\put(50,10.0){\line(0,1){80}}
\put(60,10.0){\line(0,1){80}}
\put(70,10.0){\line(0,1){80}}
\put(80,10.0){\line(0,1){80}}
\put(90,10.0){\line(0,1){80}}
\put(100,10.0){\line(0,1){80}}
\put(110,10.0){\line(0,1){80}}
\put(125,30){\footnotesize $p_2$}
\end{picture}
\begin{picture}(175,85)(450,-240)%
\put(32,82){\footnotesize $s$}
{\linethickness{.500mm}
\put(00,90){\line(1,0){170}}%
\put(00,10){\line(1,0){170}}%
\put(00,10){\line(0,1){80}}%
\put(170,10){\line(0,1){80}}}%
\put(00,70){\line(1,0){170}}
\put(00,80){\line(1,0){170}}
\put(00,60){\line(1,0){170}}
\put(00,50){\line(1,0){170}}
\put(00,40){\line(1,0){170}}
\put(00,30){\line(1,0){170}}
\put(00,20){\line(1,0){170}}
\put(10,10.0){\line(0,1){80}}
\put(20,10.0){\line(0,1){80}}
\put(30,10.0){\line(0,1){80}}
\put(40,10.0){\line(0,1){80}}
\put(50,10.0){\line(0,1){80}}
\put(60,10.0){\line(0,1){80}}
\put(70,10.0){\line(0,1){80}}
\put(80,10.0){\line(0,1){80}}
\put(90,10.0){\line(0,1){80}}
\put(100,10.0){\line(0,1){80}}
\put(110,10.0){\line(0,1){80}}
\put(120,10.0){\line(0,1){80}}
\put(130,10.0){\line(0,1){80}}
\put(140,10.0){\line(0,1){80}}
\put(150,10.0){\line(0,1){80}}
\put(160,10.0){\line(0,1){80}}%
\put(175,30){\footnotesize $p$}%
\end{picture}
\end{picture}
\ee
Massless and singleton fields on $AdS_d$
are described by  UIRs with lowest energies saturating the
unitarity bound $E_0 = E_{0}({\bf s})$. As shown in
\cite{Metsaev}, for bosonic fields
\be
\label{unitarybound}
E_0 ({\bf s})= s-p +d -2\;.
\ee
A mixed-symmetry massless higher spin bosonic  particle
with spin ${\bf s} = (s_1, ... , s_q,0, ... ,0)$ can be described
by the field
\be
\label{fr_field}
\Phi^{a_1(s_1),\,a_2(s_2),\, ...\,,\, a_{q}(s_{q})} (x) \in
B_p^{d-1,1}(s_1, \ldots , s_q )\;,
\ee
where $p$ is the height of the upper rectangular block of the Young
tableau $Y(s_1 , s_2 , \ldots, s_q
)$,\textit{ i.e.}
\be
\label{field2}
s=s_1=s_2= \cdots =s_{p} > s_{p+1}\geq
\cdots \geq s_q> 0\;, \qquad 0< p \leq q\;.
\ee
The field $\Phi^{(s_1, ... , s_q)}(x)$
generalizes the fluctuational part of the metric field in
gravitation and will be referred to as {\it metric-type} field.
(One can use either tangent or world indices since they can be
converted into each other by the background frame 1-form
$h^a$.)

The trace conditions (\ref{tr1}), (\ref{tr2}) imposed on the
metric-type field  $\Phi^{(s_1, ... , s_q)}(x)$ generalize the
Fronsdal double-tracelessness condition for totally symmetric
fields \cite{Fronsdal} to higher spin fields of any
symmetry type. Note that there are other
generalizations of Fronsdal trace conditions for mixed symmetry
fields in the literature \cite{Labastida2,Labastida,Pashnev}.
Our choice differs from some of them in that the
double-tracelessness condition is imposed on the upper rectangular
block while the rest of the Young tableau obeys the
single-tracelessness condition (the distinguished role of the
upper rectangular block is clear from (\ref{unitarybound}) and
will also be commented on later). The discrepancy between our
approach and others comes to light for non-rectangular Young
tableaux which contain at least two different blocks of length 2
or more. The simplest example is provided  by the $Y(3,2)$ tableau.

In our formulation, the flat space higher spin gauge
transformations have a structure analogous to that
proposed in \cite{Labastida}
\be
\label{gaugelaw0}
\delta\Phi^{(s_1, \,... \,, s_q)} =
\sum_{i=p}^{q}{\cal P}^{(i)}({\cal D} \xi_i^{(s_1,\, ...\, ,s_i-1, \, ...
\,, s_q)})\;,
\ee
where the gauge parameters $\xi_i^{(s_1, \,...\, ,s_i-1, \,...\, , s_q)}$
are described by various Young tableaux $Y(s_1, \,...\, ,s_i-1,\, ...\,,
s_q) $ provided that
$s_{i}-1 \geq s_{i+1}$. The gauge parameter \\ $\xi_p^{(s_1,\, ...\,
,s_1-1,s_{p+1}, \,...\,, s_q)}$
belongs to
$B^{d-1,1}_{p-1}(s_1, \,...\, ,s_1-1,s_{p+1}, \,...\,, s_q)$, while
$\xi_{k}^{(s_1,\, ...\, ,s_k-1, \,...\,, s_q)}$ for $k>p$
belongs to $B^{d-1,1}_{p}(s_1, \,...\, ,s_k-1,\, ...\,, s_q)$.
${\cal P}^{(i)}$ in (\ref{gaugelaw0}) are
projectors that involve appropriate Young symmetrizations
and take proper account of traces to project the r.h.s. to
$B^{d-1,1}_p(s_1, \ldots , s_q )$.

{}From the unitarity requirement it follows \cite{Metsaev} that
 gauge symmetries for a mixed-symmetry type field
are different in flat and $AdS_d$ backgrounds.
According to  \cite{Metsaev} the higher spin
 gauge symmetries in (\ref{gaugelaw0}) with the  parameters
$\xi_i^{(s_1, \,...\, ,s_i-1, \,...\,, s_q)}$ with $i>p$  are absent
in the $AdS_d$ background. In other words, to obtain the correct set
of $AdS_d$ gauge parameters one is allowed to cut a cell from the upper
rectangular block only. Parameters with $i>p$ appear
in the flat limit. They can play a role in the
 $AdS_d$ theory as Stueckelberg symmetry parameters, however \cite{BMV}.

\subsection{Frame-like  formulation for
mixed-symmetry bosonic massless fields}

The idea of our approach is to replace the {\it metric-type}
field $\Phi^{(s_1, ... , s_q)} (x)$ (\ref{fr_field})
by the {\it frame-type} $p$-form field
\be
\label{p-form}
\ba{l}
\omega_{(p)}{}^{a_1(s-1),\,\, ...\,, a_p(s-1),\, a_{p+1}(s_{p+1}),\,...\,,
a_{q}(s_{q})}
\\
\\
=dx^{\underline{n}_1}\wedge .... \wedge dx^{\underline{n}_p}\;
\omega_{[\underline{n}_1 ...
\underline{n}_p]}{}^{a_1(s-1),\,\, ...\,, a_p(s-1),
\, a_{p+1}(s_{p+1}),\,...\,, a_{q}(s_{q})}\,,
\ea
\ee
which takes values in the traceless tensor representation
$B^{d-1,1}_0(s-1, \, ...\,,  s-1 , s_{p+1}, \, ... \,, s_q )$
of the Lorentz group.
In other words, we cut off the last column in the upper rectangular
block of $Y(s_1, ... , s_q)$
replacing it by independent $p$-form indices.
The gauge symmetries of the $p$-form field
$\omega_{(p)}$ are required to be of the form
\be
\label{gaugelaw}
\delta \omega_{(p)}^{(s-1,\, ...\, ,  s_q)}
= {\cal D} \xi_{(p-1)}^{(s-1,\, ...\, ,  s_q)}
+ \sum_{l=p+1}^{q+1}{\cal P}^{(i)}\,(h\wedge \xi_{(p-1)}^{(s-1,\, ...\, ,
s_l+1,\, ... \,, s_q,0)})\;.
\ee
The $(p-1)$-form gauge parameter
$\xi_{(p-1)}^{(s-1,\, ...\, ,  s_q)}$,  which generalizes the
linearized diffeomorphism transformation of the frame field,
belongs to the same representation
$B^{d-1,1}_0(s-1, \, ...\,,  s-1 , s_{p+1}, \, ... \,, s_q )$
of the tangent Lorentz group
as $\omega_{(p)}$.  The $(p-1)$-form shift parameters
$\xi_{(p-1)}^{(s-1,\, ...\, , s_l+1,\, ... \,, s_q\,, s_{q+1})}\;,
p+1\leq l \leq q +1$ (with the convention that $s_{q+1}=0$) belong
to $B^{d-1,1}_0(s-1,\, ...\, , s_l+1,\, ... \,, s_q\,, s_{q+1})$
and generalize the linearized Lorentz
transformations of the frame field. They take values in the
tangent Young tableaux which differ from
that of  $\omega_{(p)}$ by one extra cell in $l^{th}$ row. This extra cell
is always contracted with the tangent index of the background
frame field $h$.
${\cal P}^{(i)}$ are projectors that take proper account of
Young symmetrizations.

The metric-type field  is expressed  in terms of the
component fields of (\ref{p-form})  as follows
\be
\label{fo}
\Phi^{a_1(s_1),\,a_2(s_2),\, ...\,,\, a_{q}(s_{q})} (x) =
\omega^{[a_1 ...a_p];\;}{}^{a_1(s-1),\,\, ...\,, a_p(s-1),
\, a_{p+1}(s_{p+1}),\,...\,, a_{q}(s_{q})}(x)\,,
\ee
\textit{i.e.} it results from symmetrization of the form indices
(converted into the tangent ones) with the tangent
indices of first $p$ rows of $\omega_{(p)}$.
{}From this formula it follows that such defined
$\Phi^{(s_1, ... , s_q)}$ belongs to $B^{d-1,1}_p(s_1, ... , s_q)$.
Indeed, the irreducible Young properties
are obvious from (\ref{fo}) since symmetrization of any
index from a lowest row with all indices of some upper row gives
zero either because of the Young properties of the tangent indices of the
component fields of (\ref{p-form}) (if a symmetrized index originates
from the tangent indices of  $\omega_{(p)}$)
or because of antisymmetry
of the form indices (if a symmetrized index is one of
the form indices of  $\omega_{(p)}$). Nonzero traces in
$\Phi^{(s_1, ... , s_q)}$ can only result from contractions of the
form indices with  tangent indices. This just gives the
conditions (\ref{tr1}), (\ref{tr2}).

The role of the gauge parameters $\xi_{(p-1)}$, which appear
in the gauge
law (\ref{gaugelaw}) without derivatives, is to compensate
redundant components of the $p$-form field (\ref{p-form})
compared to the metric-type field (\ref{fo}).
The shift symmetry in the gauge law (\ref{gaugelaw})
compensates  all  components except for the field
$\Phi^{(s_1, ... , s_q)}$.
The simplest way to see that the shift symmetry parameters do not affect
the gauge law for the field
$\Phi^{(s_1, ... , s_q)}$
is to observe that any Lorentz invariant scalar product between
$\Phi^{(s_1, ... , s_q)}$ and shift parameters gives zero
as a consequence of their Young properties, \textit{i.e.} they are
described by
different Young tableaux.

The derivative part of the gauge law
(\ref{gaugelaw}) is such that the gauge transformation of
the field $\Phi^{(s_1, ... , s_q)}$ coincides with
(\ref{gaugelaw0}) with all gauge parameters $\xi_i$ with $i>p$
absent. Thus, for the $AdS_d$ background,
the $p$-form field (\ref{p-form}) with the gauge law (\ref{gaugelaw})
describes the metric-type field $\Phi^{(s_1, ... , s_q)}$ with
the correct pattern of $AdS_d$ gauge symmetries.
It generalizes the $1$-form gauge
connection $e_{\underline{n}}{}^{a(s-1)}$ introduced in
\cite{vas_yadfiz} to describe totally symmetric spin-$s$ gauge
fields. Note that the same formalism
can be used to describe dynamics in $dS_d$.

To construct manifestly gauge invariant action one has to
introduce more fields which generalize
auxiliary and extra fields of
\cite{vas_yadfiz},\cite{V1},\cite{LV}. The idea is to associate the shift
gauge parameters of (\ref{gaugelaw}) with some new gauge fields
which generalize Lorentz connection. These will be called
auxiliary fields while the original
$p$-form (\ref{p-form}) will be referred to as physical field. The
auxiliary fields have transformation laws analogous to
(\ref{gaugelaw}) with the derivative parts containing the shift
parameters from  the gauge transformation law (\ref{gaugelaw})
of the physical $p$-form. In
addition, there will be some new shift parameters in the
transformation laws of the auxiliary fields. In their turn,
these new shift parameters require new gauge fields called
extra fields. This procedure  extends further to obtain
a full set of physical, auxiliary and extra fields
 necessary to construct curvature $(p+1)$-forms manifestly
invariant under the full set of gauge symmetries.
The analysis of the pattern of the full list of additional
gauge fields is greatly simplified by the observation
applied in \cite{d5} to the case of
totally symmetric fields that they all
result from ``dimensional reduction" of
a $p$-form gauge field
carrying an appropriate
irreducible representation of $o(d-1,2)$ (or $o(d,1)$).

\subsection{$(A)dS_d$ covariant setup}

As explained below, the appropriate set of gauge fields
is given by a $p$-form
\be
\label{adsYT}
\Omega_{(p)}{}^{A_0(r_0),\,A_1(r_1),\, ...\, ,\,A_{q}(r_q)}\;\;,
\ee
which takes values in the representation of the $AdS_d$ algebra
described by the traceless Young tableau $Y(r_0,r_1,\ldots ,r_q)$ with
\be
r_0 =r_1= \ldots =r_p = s-1\,,\qquad r_i = s_i \quad{\rm for} \quad i>p\,.
\ee
(Contraction of any two tangent indices $A$ with the
$o(d-1,2)$ invariant metric $\eta_{AB}$ gives zero.)
In other words, to describe a massless particle associated with
the vacuum energy representation (\ref{vacuum})
$B_0^{d-1,0}(\underbrace{s,\ldots , s}_p ,s_{p+1}, \ldots ,s_q )$
of $o(d-1)\subset o(d-1,2)$ we suggest to use
the $p$-form connection (\ref{adsYT}) which takes values in the
representation
$B_0^{d-1,2}(\underbrace{s-1,\ldots, s-1}_{p+1} ,s_{p+1}, \ldots, s_q )$
of the $AdS_d$ algebra $o(d-1,2)$.
The rule therefore is: to obtain the $(A)dS_d$ tensor representation
of the gauge field one cuts the shortest column and
then adds the longest row to the Young tableau of the
vacuum energy representation under consideration.
The gauge field is a $p$-form where $p$ is the height of the cut
column of the original vacuum representation.

Tensors from $B_0^{d-1,2}(\underbrace{s-1,\ldots, s-1}_{p+1} ,s_{p+1}, \ldots, s_q
)$ can be depicted as

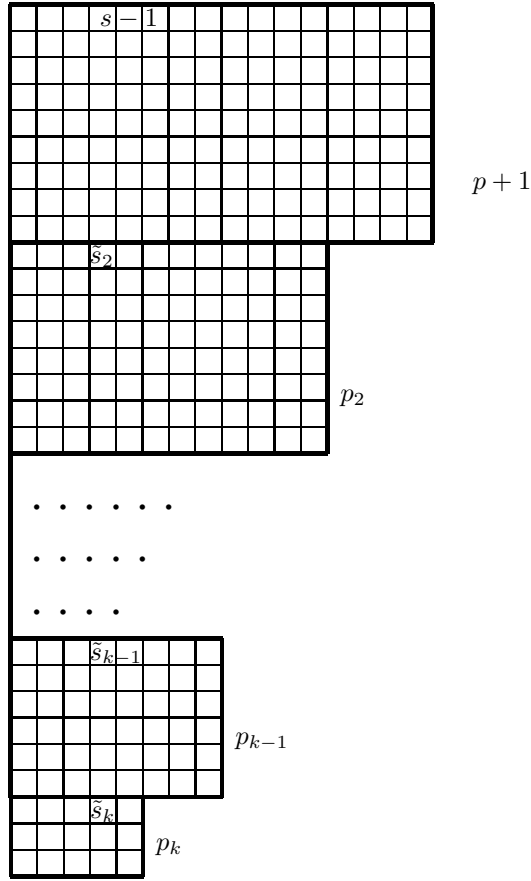
\be
\label{blockd}
\bigskip
\begin{picture}(185,350)%
\begin{picture}(109,300)%
{\linethickness{.500mm}
\put(30,32){\footnotesize $\tilde{s}_{k}$}
\put(50,10){\line(0,1){30}}
\put(00,10){\line(1,0){50}}
\put(00,10){\line(0,1){30}}
\put(00,40){\line(1,0){50}}}
\put(.0,40){\line(1,0){50}}
\put(.0,30){\line(1,0){50}}
\put(.0,20){\line(1,0){50}}
\put(10,10){\line(0,1){30}}
\put(20,10){\line(0,1){30}}
\put(30,10){\line(0,1){30}}
\put(40,10){\line(0,1){30}}
\put(55,20){\footnotesize $p_{k}$}
\end{picture}
\begin{picture}(125,90)(113,-30)%
{\linethickness{.500mm}
\put(30,62){\footnotesize $\tilde{s}_{k-1}$}
\put(80,10){\line(0,1){60}}
\put(00,10){\line(1,0){80}}
\put(00,10){\line(0,1){60}}
\put(00,70){\line(1,0){80}}}
\put(.0,60){\line(1,0){80}}
\put(.0,50){\line(1,0){80}}
\put(.0,40){\line(1,0){80}}
\put(.0,30){\line(1,0){80}}
\put(.0,20){\line(1,0){80}}
\put(10,10){\line(0,1){60}}
\put(20,10){\line(0,1){60}}
\put(30,10){\line(0,1){60}}
\put(40,10){\line(0,1){60}}
\put(50,10){\line(0,1){60}}
\put(60,10){\line(0,1){60}}
\put(70,10){\line(0,1){60}}
\put(85,30){\footnotesize $p_{k-1}$}
\end{picture}
\begin{picture}(75,75)(242,-90)%
{\linethickness{.5mm}
\put(00,10){\line(0,1){70}}}
\put(10,20){\circle*{2}}
\put(20,20){\circle*{2}}
\put(30,20){\circle*{2}}
\put(40,20){\circle*{2}}
\put(50,40){\circle*{2}}
\put(40,40){\circle*{2}}
\put(30,40){\circle*{2}}
\put(10,40){\circle*{2}}
\put(20,40){\circle*{2}}
\put(60,60){\circle*{2}}
\put(50,60){\circle*{2}}
\put(40,60){\circle*{2}}
\put(30,60){\circle*{2}}
\put(10,60){\circle*{2}}
\put(20,60){\circle*{2}}
\end{picture}
\begin{picture}(125,95)(321,-160)%
\put(30,82){\footnotesize $\tilde{s}_2$}%
{\linethickness{.500mm}
\put(120,10){\line(0,1){80}}
\put(00,90){\line(1,0){120}}
\put(00,10){\line(1,0){120}}
\put(00,10){\line(0,1){80}}}
\put(00,70){\line(1,0){120}}
\put(00,80){\line(1,0){120}}
\put(00,60){\line(1,0){120}}
\put(00,50){\line(1,0){120}}
\put(00,40){\line(1,0){120}}
\put(00,30){\line(1,0){120}}
\put(00,20){\line(1,0){120}}
\put(10,10.0){\line(0,1){80}}
\put(20,10.0){\line(0,1){80}}
\put(30,10.0){\line(0,1){80}}
\put(40,10.0){\line(0,1){80}}
\put(50,10.0){\line(0,1){80}}
\put(60,10.0){\line(0,1){80}}
\put(70,10.0){\line(0,1){80}}
\put(80,10.0){\line(0,1){80}}
\put(90,10.0){\line(0,1){80}}
\put(100,10.0){\line(0,1){80}}
\put(110,10.0){\line(0,1){80}}
\put(125,30){\footnotesize $p_2$}
\end{picture}
\begin{picture}(175,85)(450,-240)%
\put(34,92){\footnotesize $s-1$}
{\linethickness{.500mm}
\put(00,100){\line(1,0){160}}
\put(00,10){\line(1,0){160}}
\put(00,10){\line(0,1){90}}
\put(160,10){\line(0,1){90}}
}
\put(00,90){\line(1,0){160}}
\put(00,70){\line(1,0){160}}
\put(00,80){\line(1,0){160}}
\put(00,60){\line(1,0){160}}
\put(00,50){\line(1,0){160}}
\put(00,40){\line(1,0){160}}
\put(00,30){\line(1,0){160}}
\put(00,20){\line(1,0){160}}
\put(10,10.0){\line(0,1){90}}
\put(20,10.0){\line(0,1){90}}
\put(30,10.0){\line(0,1){90}}
\put(40,10.0){\line(0,1){90}}
\put(50,10.0){\line(0,1){90}}
\put(60,10.0){\line(0,1){90}}
\put(70,10.0){\line(0,1){90}}
\put(80,10.0){\line(0,1){90}}
\put(90,10.0){\line(0,1){90}}
\put(100,10.0){\line(0,1){90}}
\put(110,10.0){\line(0,1){90}}
\put(120,10.0){\line(0,1){90}}
\put(130,10.0){\line(0,1){90}}
\put(140,10.0){\line(0,1){90}}
\put(150,10.0){\line(0,1){90}}
\put(160,10.0){\line(0,1){90}}
\put(175,30){\footnotesize $p+1$}%
\end{picture}
\end{picture}
\ee
The set of various $p$-form
Lorentz-covariant gauge fields, including physical, auxiliary
and extra fields, associated with a particular mixed-symmetry
representation of the $AdS_d$ algebra, is in the one-to-one
correspondence with the set of irreducible representations of the Lorentz
subalgebra $o(d-1,1)\subset o(d-1,2)$, contained in
 \newline $B_0^{d-1,2}(\underbrace{s-1,\ldots, s-1}_{p+1} ,s_{p+1}, \ldots, s_q )$.
In practice, Lorentz-covariant component fields are identified with various
independent traceless $V$-transversal components in the original
$o(d-1,2)$ traceless Young tableau. For the particular case
of gravitation  it works as in Eq.(\ref{defEOmega}).
The decomposition of an arbitrary mixed-symmetry  field yields
the set of $p$-form  Lorentz-covariant tensor
fields represented by the following Young tableaux:

\be
\label{lorblock}
\begin{picture}(165,350)%
\begin{picture}(112,300)%
{\linethickness{.5mm}
\put(31,32){\footnotesize $\tilde{s}_{k}$}
\put(50,20){\line(0,1){20}}  
\put(00,10){\line(1,0){30}}  
\put(00,20){\line(1,0){50}}  
\put(30,10){\line(0,1){10}}  
\put(00,10){\line(0,1){30}}  
\put(00,40){\line(1,0){50}}  
}
\put(.0,40){\line(1,0){50}}
\put(.0,30){\line(1,0){50}}  
\put(.0,20){\line(1,0){50}}
\put(20,10){\line(1,0){30}}  
\put(50,10){\line(0,1){10}}  
\put(10,10){\line(0,1){30}}  
\put(20,10){\line(0,1){30}}  
\put(30,20){\line(0,1){20}}  
\put(40,10){\line(0,1){30}}  
\put(12,12){\footnotesize $t_{k}$}
\put(53,23){\footnotesize $p_{k}-1$}
\put(43,13){\scriptsize o}
\put(33,13){\scriptsize o}
\end{picture}
\begin{picture}(125,90)(116,-30)%
{\linethickness{.500mm}
\put(31,62){\footnotesize $\tilde{s}_{k-1}$}
\put(80,20){\line(0,1){50}}
\put(00,10){\line(1,0){60}}
\put(60,10){\line(0,1){10}}
\put(0,20){\line(1,0){80}}
\put(00,10){\line(0,1){60}}
\put(00,70){\line(1,0){80}}
}
\put(80,10){\line(0,1){60}}   
\put(20,10){\line(1,0){60}}   
\put(.0,60){\line(1,0){80}}
\put(.0,50){\line(1,0){80}}
\put(.0,40){\line(1,0){80}}
\put(.0,30){\line(1,0){80}}
\put(.0,20){\line(1,0){80}}
\put(10,10){\line(0,1){60}}
\put(20,10){\line(0,1){60}}
\put(30,10){\line(0,1){60}}
\put(40,10){\line(0,1){60}}
\put(50,10){\line(0,1){60}}
\put(60,20){\line(0,1){50}}
\put(70,10){\line(0,1){60}}
\put(11,12){\footnotesize $t_{k-1}$}
\put(85,30){\footnotesize $p_{k-1}-1$}
\put(73,13){\scriptsize o}
\put(63,13){\scriptsize o}
\end{picture}
\begin{picture}(75,75)(245,-90)%
{\linethickness{.5mm} \put(00,10){\line(0,1){70}}}
\put(10,20){\circle*{2}} \put(20,20){\circle*{2}}
\put(30,20){\circle*{2}} \put(40,20){\circle*{2}}
\put(50,40){\circle*{2}} \put(40,40){\circle*{2}}
\put(30,40){\circle*{2}} \put(10,40){\circle*{2}}
\put(20,40){\circle*{2}} \put(60,60){\circle*{2}}
\put(50,60){\circle*{2}} \put(40,60){\circle*{2}}
\put(30,60){\circle*{2}} \put(10,60){\circle*{2}}
\put(20,60){\circle*{2}} \end{picture}
\begin{picture}(125,95)(324,-160)%
\put(61,82){\footnotesize $\tilde{s}_2$}%
{\linethickness{.5mm}
\put(120,20){\line(0,1){70}}
\put(00,90){\line(1,0){120}}
\put(00,10){\line(1,0){90}}
\put(90,10){\line(0,1){10}}
\put(0,20){\line(1,0){120}}
\put(00,10){\line(0,1){80}}
}
\put(00,10){\line(1,0){120}}   
\put(120,10){\line(0,1){80}}  
\put(00,70){\line(1,0){120}}
\put(00,80){\line(1,0){120}}
\put(00,60){\line(1,0){120}}
\put(00,50){\line(1,0){120}}
\put(00,40){\line(1,0){120}}
\put(00,30){\line(1,0){120}}
\put(00,20){\line(1,0){120}}
\put(10,10.0){\line(0,1){80}}
\put(20,10.0){\line(0,1){80}}
\put(30,10.0){\line(0,1){80}}
\put(40,10.0){\line(0,1){80}}
\put(50,10.0){\line(0,1){80}}
\put(60,10.0){\line(0,1){80}}
\put(70,10.0){\line(0,1){80}}
\put(80,10.0){\line(0,1){80}}
\put(90,20){\line(0,1){70}}
\put(100,10.0){\line(0,1){80}}
\put(110,10.0){\line(0,1){80}}
\put(32,12){\footnotesize $t_2$}
\put(125,40){\footnotesize $p_2-1$}
\put(93,13){\scriptsize o}
\put(103,13){\scriptsize o}
\put(113,13){\scriptsize o}
\end{picture}
\begin{picture}(175,85)(453,-240)%
\put(63.4,92){\footnotesize $s-1$}
{\linethickness{.5mm}
\put(00,100){\line(1,0){160}}%
\put(00,10){\line(1,0){140}}%
\put(140,10){\line(0,1){10}}%
\put(0,20){\line(1,0){160}}%
\put(00,10){\line(0,1){90}}%
\put(160,20){\line(0,1){80}}%
}
\put(30,10){\line(1,0){130}}    
\put(00,90){\line(1,0){160}}
\put(00,70){\line(1,0){160}}
\put(00,80){\line(1,0){160}}
\put(00,60){\line(1,0){160}}
\put(00,50){\line(1,0){160}}
\put(00,40){\line(1,0){160}}
\put(00,30){\line(1,0){160}}
\put(10,10.0){\line(0,1){90}}
\put(20,10.0){\line(0,1){90}}
\put(30,10.0){\line(0,1){90}}
\put(40,10.0){\line(0,1){90}}
\put(50,10.0){\line(0,1){90}}
\put(60,10.0){\line(0,1){90}}
\put(70,10.0){\line(0,1){90}}
\put(80,10.0){\line(0,1){90}}
\put(90,10.0){\line(0,1){90}}
\put(100,10.0){\line(0,1){90}}
\put(110,10.0){\line(0,1){90}}
\put(120,10.0){\line(0,1){90}}
\put(130,10.0){\line(0,1){90}}
\put(140,20.0){\line(0,1){80}}
\put(150,10.0){\line(0,1){90}}
\put(160,10.0){\line(0,1){90}}%
\put(31,12){\footnotesize $t_1$}
\put(175,40){\footnotesize $p$}%
\put(143,13){\scriptsize o}
\put(153,13){\scriptsize o}
\end{picture}
\end{picture}
\ee
where
\be
\label{t}
\tilde{s}_{i+1}\leq  t_i \leq \tilde{s}_i \;,
\ee
(with the convention that $\tilde{s}_1=s-1$ and $\tilde{s}_{k+1}=0$).
The corresponding Lorentz-covariant tensors result from contractions of
some of the
indices of the original $o(d-1,2)$ tensor with the compensator $V^A$
along with projecting out the $V^A$ transversal components with respect
to the rest of indices. In the tableau (\ref{lorblock}) the indices
contracted with the compensator are denoted as
$\begin{picture}(10,10)(-1,5)
{\linethickness{0.210mm}
\put(00,12){\line(1,0){7}} 
\put(00,05){\line(1,0){7}} 
\put(00,05){\line(0,1){7}} 
\put(07,05){\line(0,1){7}} 
}
\put(1.6,06.5){\scriptsize o}
\end{picture}$.
They disappear from the resulting Lorentz tableau
drown in bold because the compensator is, by definition,
Lorentz invariant. Clearly the result is equivalent to the
dimensional reduction of an irreducible  tensor to one lower
dimension\footnote{The interpretation of the picture (\ref{lorblock})
is somewhat schematic in that respect that the straightforward
dimensional
reduction of an irreducible Young tableau by contracting some of the
indices with the compensator does not generically produce an
irreducible lower-dimensional (i.e., Lorentz in our case)
Young tableau. Nevertheless, one can see that the list of the
resulting irreducible components  is correctly
reproduced by the bold Young tableaux in (\ref{lorblock}).}.
Note that in the list of resulting Lorentz tableaux
no two contractions with the compensator hit the same column, i.e.,
no two cells
$\begin{picture}(10,10)(-1,5)
{\linethickness{0.210mm}
\put(00,12){\line(1,0){7}} 
\put(00,05){\line(1,0){7}} 
\put(00,05){\line(0,1){7}} 
\put(07,05){\line(0,1){7}} 
}
\put(1.6,06.5){\scriptsize o}
\end{picture}$ are situated one under another.
This is because the product of two
compensators $V^AV^B$ is a symmetric tensor.

To summarize, the decomposition of the $(A)dS_d$ $p$-form
gauge field with tangent indices given by the traceless
$(o(d-1,2))o(d,1)$  Young tableau (\ref{blockd}) into a
set of Lorentz-covariant $p$-form fields is
\be
\Omega_{(p)} \rightarrow \bigoplus_{(t_1, ..., \,t_k)}
\omega_{(p)}{}^{(t_1, ..., \,t_k)}\;,
\ee
where the fields $\omega_{(p)}{}^{(t_1, ..., \,t_k)}$ parameterized
by the integers $t_i$ (\ref{t}) have tangent
indices given by various traceless $o(d-1,1)$ Young tableaux
(\ref{lorblock}), (\ref{t}).

The dynamical interpretation of different Lorentz-covariant
fields is as follows:

\begin{itemize}
\item
The physical field corresponds to
the tableau (\ref{lorblock}) with $t_i = \tilde{s}_{i+1}$ for all $i$,
which is equivalent to cut off  the upper row in the
Young tableau (\ref{blockd}). This means that the physical field
is identified with the maximally $V$-tangential component of the $(A)dS$
field (\ref{adsYT}), associated with the contraction of its $s-1$
indices with $V^A$, \textit{i.e.}
\be
\omega_{(p)}{}^{a_1(s-1),\,\, ...\,, a_p(s-1),\, a_{p+1}(s_{p+1}),\,...\,,
a_{q}(s_{q})}
=\underbrace{V_{A_0} \ldots V_{A_0}}_{s-1}
\Omega_{(p)}{}^{A_0(s-1),\,a_1(s-1),\, ...\, ,\,a_{q}(s_q)}\,.
\ee
Note that contraction of any $s$ indices of
$\Omega_{(p)}{}^{A_0(s-1),\, ...\, ,\,A_{q}(s_q)}$
with $V^A$ gives zero because of the Young properties of
$\Omega_{(p)}{}^{A_0(s-1),\, ...\, ,\,A_{q}(s_q)}$.

\item
The auxiliary fields have tangent Lorentz tableaux
which differ from the physical field
by one cell, {\it i.e.} they correspond to tableaux (\ref{lorblock}) with
$  t_j =  \tilde{s}_{j+1}+1$ for some particular $j$, while $t_i =
\tilde{s}_{i+1}$ for
all other $i$. There are $k$ different auxiliary fields for a Young
tableau composed of $k$
blocks (\ref{blockd}).
\item
The class of extra fields includes all the rest
Lorentz tableaux (\ref{lorblock}) having two or more
additional cells compared to the physical field tableau.

\end{itemize}

The linearized higher spin curvature $(p+1)$-form  associated with the
gauge
$p$-form field (\ref{adsYT}) - (\ref{blockd}) is
\be
\label{adscurv}
\ba{c}
R_{(p+1)}{}^{A_0(s-1),\, ...\, ,\,A_{q}(s_q)} = D_0
\Omega_{(p)}{}^{A_0(s-1),\, ...\, ,\,A_{q}(s_q)}\;,
\ea
\ee
where the $(o(d-1,2))o(d,1)$ covariant derivative $D_0$ is defined
 according to (\ref{backderaction}) with respect to some background
$(A)dS_d$  connection $W^{AB}$ (\ref{zerocurv}).

The curvature $(p+1)$-form is manifestly invariant
\be
\label{adsgaugetr0}
\delta R_{(p+1)}{}^{A_0(s-1),\, ...\, ,\,A_{q}(s_q)} = 0\;
\ee
under the gauge transformations
\be
\label{adsgaugetr}
\delta\Omega_{(p)}{}^{A_0(s-1),\, ...\, ,\,A_{q}(s_q)} = D_0
\xi_{(p-1)}{}^{A_0(s-1),\, ...\, ,\,A_{q}(s_q)}\;
\ee
with the $(p-1)$-form gauge parameter
$\xi_{(p-1)}{}^{A_0(s-1),\, ...\, ,\,A_{q}(s_q)}$
and satisfies Bianchi identities
\be
D_0 R_{(p+1)}{}^{A_0(s-1),\, ...\, ,\,A_{q}(s_q)} = 0\;
\ee
as a consequence of the
zero-curvature condition $D_0^2=0$ (\ref{zerocurv}).
Another consequence of the zero-curvature condition
is that the gauge transformations (\ref{adsgaugetr}) are reducible.
There exists the set of level-$(l+2)$ ($0\leq l \leq p-2$) gauge parameters
and gauge transformations of the form
\be
\delta\xi_{(p-l-1)}{}^{A_0(s-1),\, ...\, ,\,A_{q}(s_q)} = D_0
\xi_{(p-l-2)}{}^{A_0(s-1),\, ...\, ,\,A_{q}(s_q)}\;.
\ee

The gauge transformation  law (\ref{adsgaugetr}) gives precise form of
(\ref{gaugelaw}) along with gauge transformations for  all auxiliary
and extra fields.

\subsection{Fermionic mixed-symmetry massless fields }

Formulation of  mixed-symmetry fermionic massless fields is analogous.
Consider a fermionic field which describes upon quantization
the unitary module of the $AdS_d$ symmetry group $o(d-1,2)$ induced
from the vacuum module of its maximal compact subgroup
$o(2)\oplus o(d-1)$, characterized
by some energy $E_0$ and ``spin'' ${\bf  s} = (h_1, \ldots, h_q, 1/2, ..., 1/2)$
with $h_1 \geq h_2\geq \ldots \geq h_q > 1/2$,
where all $2h_i$ are odd and $q\leq \nu = [\frac{d-1}{2}]$.
In terms of the Young tableau associated with the corresponding
spinor-tensor representation of $o(d-1)$,
$s_i = (h_i-1/2)$ is the length of its $i^{th}$ row. The vacuum
energy $E_0$ of massless fermionic fields is \cite{Metsaev}
\be
\label{unitarybound_f}
E_0 = s_1 - p +d-\frac{3}{2}\;,
\ee
where $p$ is the height of the upper rectangular block.

Introduce a Lorentz-covariant \textit{spinor-tensor} field
\be
\label{fermion1}
\psi^{\alpha\;|a_1(s_1),\,a_2(s_2),\, ...\,,\, a_{q}(s_{q})}\;,
\ee
which is  symmetric in each group of indices
$a_i (s_i )$ and satisfies the
Young symmetry conditions associated with the Young tableau
$Y(s_1 , s_2 , \ldots, s_q )$.  ($\alpha$  is
Lorentz spinor index.)

Let the spinor-tensor (\ref{fermion1}) satisfy  the conditions
\be
\label{ftr1}
(\gamma_{a_i}\gamma_{a_i}\gamma_{a_i})^{\alpha}{}_{\beta}\,\psi^{\beta\;|a_1(s_1),\,a_2(s_2),\,
...\,,\, a_{q}(s_{q})}=0\;, \qquad 0<  i \leq m\,,
\ee
and
\be
\label{ftr2}
\gamma_{a_i}{}^\alpha{}_\beta\,
\psi^{\beta\;|a_1(s_1),\,a_2(s_2),\, ...\,,\, a_{q}(s_{q})}=0\;, \qquad m<
i \leq q\,,
\ee
where $\gamma_{a}$ are Dirac  matrices,
$\{\gamma_a,\,\gamma_b\}=2\eta_{ab}$,
and $m$ is some non-negative integer. The linear space
of $o(d-1,1)$ spinor-tensors (\ref{fermion1}) which have
the Young properties of the type $Y(s_1,...,s_q )$ and satisfy
the conditions (\ref{ftr1}), (\ref{ftr2}) will be denoted
$F^{d-1,1}_m(s_1, \ldots , s_q )$ (respectively,
$F^{p,r}_m(s_1, \ldots , s_q )$ for $o(p,r)$).
It follows from the condition (\ref{ftr1}) that contraction of any two
pairs of indices from any of the first
$m$ rows of the Young tableau gives zero. Also, it follows
from (\ref{ftr2}) that contraction of
any pair of indices from the last $q - m$ rows gives zero.

To describe $AdS_d$ dynamics of a
spin-${\bf s}$ massless fermion, introduce a Lorentz-covariant
\textit{spinor-tensor} field
\be
\label{fermion11}
\psi^{\alpha\;|a_1(s_1),\,a_2(s_2),\, ...\,,\, a_{q}(s_{q})}(x)\in
F^{d-1,1}_p(s_1, \ldots , s_q )\;,
\ee
where $p$ is the height of the upper rectangular block
of the Young tableau $Y(s_1 , s_2 , \ldots s_q)$,\textit{ i.e.}
\be
\label{ffield2}
s=s_1=s_2= \cdots =s_{p} > s_{p+1}\geq
\cdots \geq s_q> 0\;, \qquad 0< p \leq q\;.
\ee

Analogously to the case of bosonic fields,
the metric-type field (\ref{fermion11}) is replaced by the
$p$-form spinor-tensor field\footnote{The
analogous construction was exploited in \cite{alk1,Alkalaev} to
describe symmetric massless fermions in $AdS_5$.}
\be
\label{psi}
\Omega_{(p)}{}^{\hat{\alpha}\;|A_0 (s-1)\,,\, ... \,,\, A_q(s_q)}\;,
\ee
which take values in the representation
$F_0^{d-1,2}(\underbrace{s-1,\ldots, s-1}_{p+1} ,s_{p+1}, \ldots, s_q )$
  of $o(d-1,2)$.
Here $\hat{\alpha}$ is some irreducible  $o(d-1,2)$ spinor index,
\textit{i.e.} it is Majorana or Weyl or both, whenever possible.
The Lorentz spinor index $\ga$ is then identified with
the $o(d-1,2)$ spinor index $\hat{\ga}$, which means that it is not
necessarily irreducible (e.g., chiral) with respect to $o(d-1,1)$.

The  background covariant   derivative acts on an arbitrary
$AdS_d$ spinor-tensor in the standard way
\be
\ba{c}
D_0\Upsilon{}^{\hat{\alpha}\;|A_1 (s_1)\,,\, ... \,,\, A_m(s_m)}
\\
\\
\dps= d\,\Upsilon{}^{\hat{\alpha}\;|A_1 (s_1)\,,\, ... \,,\, A_m(s_m)}\;
+ \frac{1}{2}\,W_{BC}\,(\sigma^{BC})^{\hat{\alpha}}{}_{\hat{\beta}}\;
\,\wedge\,\Upsilon{}^{\hat{\beta}\;|A_1 (s_1)\,,\, ... \,,\, A_m(s_m)}
\\
\\
+s_1 \,W^{A_1}{}_C\wedge \Upsilon{}^{\hat{\alpha}\;|CA_1(s_1-1)\,,\, ...
\,,\, A_m(s_m)} + \ldots +
s_m \, W^{A_m}{}_C\wedge \Upsilon{}^{\hat{\alpha}\;|A_1(s_1)\,,\, ...
\,,\, CA_m(s_m-1)}\;,
\ea
\ee
where the background connection $W^{AB}$ satisfies the zero-curvature
condition (\ref{zerocurv}) and
$$
\sigma^{AB}= \frac{1}{4}
[\gamma^A,\,\gamma^B]\,,\qquad
\{\gamma_A,\,\gamma_B\}=2\eta_{AB}.
$$

The fermionic curvature
\be
\label{adscurvf}
R_{(p+1)}{}^{\hat{\alpha}\;|A_0 (s-1)\,,\, ... \,,\, A_q(s_q)}=
D_0\Omega_{(p)}{}^{\hat{\alpha}\;|A_0 (s-1)\,,\, ... \,,\, A_q(s_q)}\;
\ee
is invariant
\be
\label{adsgaugetr0f}
\delta R_{(p+1)}{}^{\hat{\alpha}\;|A_0(s-1),\, ...\, ,\,A_{q}(s_q)} = 0\;
\ee
under the gauge transformations
\be
\label{adsgaugetrf}
\delta\Omega_{(p)}{}^{\hat{\alpha}\;|A_0 (s-1)\,,\, ... \,,\, A_q(s_q)}\;
=D_0 \xi_{(p-1)}{}^{\hat{\alpha}\;|A_0 (s-1)\,,\, ... \,,\, A_q(s_q)}\;.
\ee

The physical field is
\be
\omega_{(p)}{}^{\ga\;|a_1(s-1),\,\, ...\,, a_p(s-1),\,
a_{p+1}(s_{p+1}),\,...\,, a_{q}(s_{q})}
=\underbrace{V_{A_0} \ldots V_{A_0}}_{s-1}
\Omega_{(p)}{}^{\hat{\ga}\;|A_0(s-1),\,a_1(s-1),\, ...\, ,\,a_{q}(s_q)}\,.
\ee
In the fermionic case all other components in
\be
\Omega_{(p)}{}^{\hat{\alpha}\;|A_0 (s-1)\,,\, ... \,,\, A_q(s_q)}\;
\ee
will be called extra fields.
The metric-type component is defined in terms of physical field
analogously to the bosonic case (\ref{fo})
\be
\label{fofe}
\psi^{\ga\,|a_1(s_1),\,a_2(s_2),\, ...\,,\, a_{q}(s_{q})} (x) =
\omega^{\ga\,|\,[a_1 ...a_p];\,a_1(s-1),\,\, ...\,, a_p(s-1),
\, a_{p+1}(s_{p+1}),\,...\,, a_{q}(s_{q})}(x)\,.
\ee
As a result, it satisfies the conditions
(\ref{ftr1}) and (\ref{ftr2}) and belongs to $F^{d-1,1}_p(s_1, \ldots ,
s_q )$.

\subsection{Action and Weyl tensors}

Having constructed the  gauge invariant
linearized curvatures (\ref{adscurv}) and (\ref{adscurvf}) one can
look for a free action functional in the form \cite{LV}
\be
\label{action}
{\cal S}_2 = \int_{{\cal M}^d}\; \alpha^{\cdots} (V)
\;\underbrace{E^{\cdots}\wedge ... \wedge E^{\cdots}}_{d-2p-2}\wedge\;
R_{(p+1)}^{\cdots} \wedge \,R_{(p+1)}^{\cdots}\;.
\ee
 Here $E$ is the background frame
field and $\alpha^{\cdots} (V)$ are some
coefficients which parameterize various types of index
contractions between curvatures, frame fields and compensators.
Any such action is gauge invariant with respect to the
full set of gauge transformations because of (\ref{adsgaugetr0}),
(\ref{adsgaugetr}) and (\ref{adsgaugetr0f}), (\ref{adsgaugetrf}).

The coefficients have to be determined by imposing {\it the extra
field decoupling condition} which, effectively, requires the
action to be free of higher derivatives of the physical field.
Actually, in the fermionic case, the higher spin equations should
be first-order that will be true if the variation of ${\cal S}_2 $
with respect to all extra fields is demanded to identically vanish
\cite{vf}. In the bosonic case, the higher spin equations are of
second-order. As the auxiliary fields are expressed by virtue of
their equations of motion in terms of first derivatives of the
metric-type field $\Phi^{(s_1, ... , s_q)}$ modulo pure gauge
parts, the bosonic equations of motion will be of second-order
once the variation of ${\cal S}_2 $ with respect to extra fields
vanishes identically \cite{LV}. As soon as the coefficients that
guarantee independence of
 extra fields are found, the resulting action will,
by construction, be  invariant under correct higher spin gauge
transformations and give rise to invariant  differential equations
for the metric-type field $\Phi^{a_1(s_1),\,a_2(s_2),\, ...\,,\,
a_{q}(s_{q})} (x)$ in the bosonic case and
$\psi^{\ga\;|a_1(s_1),\,a_2(s_2),\, ...\,,\, a_{q}(s_{q})} (x)$ in
the fermionic case. Realization of this program for a general
massless field will be given elsewhere. In Section 3 we illustrate
how it works for some simple examples.

Although extra fields do not contribute to the free higher spin action,
they play a role at the interaction level. It is therefore
necessary to express auxiliary and extra fields in terms of
derivatives of the physical fields by appropriate constraints.
The strategy that proved to be most appropriate for the case
of symmetric fields \cite{V1,LV,vf} is to impose constrains in
terms of linearized higher spin curvatures setting to zero as
many of their components as possible
to express algebraically auxiliary and extra fields
in terms of derivatives of the physical fields. These
constraints generalize the zero-torsion constraint in gravity
which expresses Lorentz connection in terms of
first derivatives of the physical field identified with the
frame 1-form.

The analysis of constraints for the case of generic mixed symmetry
massless fields is important in several  respects. In particular,
it provides a starting point towards the unfolded formulation of
nonlinear higher spin dynamics in the form of appropriate
covariant  constancy conditions. For the case of totally symmetric
massless fields it is known \cite{LV,vf} that the constraints for
auxiliary and extra fields along with the field equations and
Bianchi identities allow one to set equal to zero most of the
components of the linearized higher spin curvatures. By analogy
with gravity, the non-zero
components are called higher spin Weyl tensors. As imposed
constraints express all fields in terms of derivatives of the
physical higher spin field, the generalized Weyl tensors turn out
to be expressed in terms of derivatives of the physical field and,
being components of the higher spin curvatures, remain invariant
under the higher spin gauge transformations. As a result,
generalized Weyl tensors parameterize those gauge invariant
combinations of derivatives of the physical fields which remain
nonzero on-mass-shell. The full analysis of their structure, as
well as of the  structure of constraints, is beyond the scope of
this paper. Here we only would like to note that our construction
naturally gives rise to the Weyl tensors analogous to the gauge
invariant mixed symmetry higher spin curvatures found by Medeiros
and Hull in \cite{hull} within non-local
formulation of higher spin dynamics in Minkowski space.

Since an irreducible $AdS_d$ system decomposes in the flat limit
into a set of independent mixed-symmetry Minkowski higher spin
fields, it is natural to conjecture that the Weyl tensors
associated with any irreducible flat space subsystem should
appear. {}From the analysis of \cite{BMV,hull} it follows that the
set of flat space generalized Weyl tensors associated with the
$AdS_d$ mixed-symmetry field, described in terms of a
$p$-form connection taking values in the $AdS_d$ Young tableau
(\ref{blockd}), is given by the set of various Young tableaux of the form

\be
\label{adweyl}
\begin{picture}(185,350)%
\begin{picture}(112,300)%
{\linethickness{.5mm}
\put(30,32){\footnotesize $\tilde{s}_{k}$}
\put(50,20){\line(0,1){20}} \put(00,10){\line(1,0){30}}
\put(00,20){\line(1,0){50}} \put(30,10){\line(0,1){10}}
\put(00,10){\line(0,1){30}} \put(00,40){\line(1,0){50}}}
\put(.0,40){\line(1,0){50}} \put(.0,30){\line(1,0){50}}
\put(.0,20){\line(1,0){50}} \put(10,10){\line(0,1){30}}
\put(20,10){\line(0,1){30}} \put(30,20){\line(0,1){20}}
\put(40,20){\line(0,1){20}}
\put(10,13){\footnotesize $n_{k}$}
\put(53,23){\footnotesize $p_{k}-1$}
\end{picture}
\begin{picture}(125,90)(116,-30)%
{\linethickness{.500mm}
\put(30,62){\footnotesize $\tilde{s}_{k-1}$}
\put(80,20){\line(0,1){50}}
\put(00,10){\line(1,0){60}} \put(60,10){\line(0,1){10}}
\put(0,20){\line(1,0){80}} \put(00,10){\line(0,1){60}}
\put(00,70){\line(1,0){80}}} \put(.0,60){\line(1,0){80}}
\put(.0,50){\line(1,0){80}} \put(.0,40){\line(1,0){80}}
\put(.0,30){\line(1,0){80}} \put(.0,20){\line(1,0){80}}
\put(10,10){\line(0,1){60}} \put(20,10){\line(0,1){60}}
\put(30,10){\line(0,1){60}} \put(40,10){\line(0,1){60}}
\put(50,10){\line(0,1){60}} \put(60,20){\line(0,1){50}}
\put(70,20){\line(0,1){50}}
\put(10,13){\footnotesize $n_{k-1}$}
\put(85,30){\footnotesize $p_{k-1}-1$} %
\end{picture}
\begin{picture}(75,75)(245,-90)%
{\linethickness{.5mm} \put(00,10){\line(0,1){70}}}
\put(10,20){\circle*{2}} \put(20,20){\circle*{2}}
\put(30,20){\circle*{2}} \put(40,20){\circle*{2}}
\put(50,40){\circle*{2}} \put(40,40){\circle*{2}}
\put(30,40){\circle*{2}} \put(10,40){\circle*{2}}
\put(20,40){\circle*{2}} \put(60,60){\circle*{2}}
\put(50,60){\circle*{2}} \put(40,60){\circle*{2}}
\put(30,60){\circle*{2}} \put(10,60){\circle*{2}}
\put(20,60){\circle*{2}} \end{picture}
\begin{picture}(125,95)(324,-160)%
\put(30,82){\footnotesize $\tilde{s}_2$}%
{\linethickness{.5mm}
\put(120,20){\line(0,1){70}}
\put(00,90){\line(1,0){120}} \put(00,10){\line(1,0){90}}
\put(90,10){\line(0,1){10}} \put(0,20){\line(1,0){120}}
\put(00,10){\line(0,1){80}}} \put(00,70){\line(1,0){120}}
\put(00,80){\line(1,0){120}} \put(00,60){\line(1,0){120}}
\put(00,50){\line(1,0){120}} \put(00,40){\line(1,0){120}}
\put(00,30){\line(1,0){120}} \put(00,20){\line(1,0){120}}
\put(10,10.0){\line(0,1){80}} \put(20,10.0){\line(0,1){80}}
\put(30,10.0){\line(0,1){80}} \put(40,10.0){\line(0,1){80}}
\put(50,10.0){\line(0,1){80}} \put(60,10.0){\line(0,1){80}}
\put(70,10.0){\line(0,1){80}} \put(80,10.0){\line(0,1){80}}
\put(90,20){\line(0,1){70}} \put(100,20.0){\line(0,1){70}}
\put(110,20.0){\line(0,1){70}}
\put(40,13){\footnotesize $n_2$}
\put(125,40){\footnotesize $p_2-1$}
\end{picture}
\begin{picture}(175,85)(453,-240)%
\put(34,92){\footnotesize $s$}
{\linethickness{.500mm}
\put(00,100){\line(1,0){170}}%
\put(00,10){\line(1,0){170}}%
\put(00,10){\line(0,1){90}}%
\put(170,10){\line(0,1){90}}}%
\put(00,90){\line(1,0){170}}
\put(00,80){\line(1,0){170}}
\put(00,70){\line(1,0){170}}
\put(00,60){\line(1,0){170}}
\put(00,50){\line(1,0){170}}
\put(00,40){\line(1,0){170}}
\put(00,30){\line(1,0){170}}
\put(00,20){\line(1,0){170}}
\put(10,10.0){\line(0,1){90}}
\put(20,10.0){\line(0,1){90}}
\put(30,10.0){\line(0,1){90}}
\put(40,10.0){\line(0,1){90}}
\put(50,10.0){\line(0,1){90}}
\put(60,10.0){\line(0,1){90}}
\put(70,10.0){\line(0,1){90}}
\put(80,10.0){\line(0,1){90}}
\put(90,10.0){\line(0,1){90}}
\put(100,10.0){\line(0,1){90}}
\put(110,10.0){\line(0,1){90}}
\put(120,10.0){\line(0,1){90}}
\put(130,10.0){\line(0,1){90}}
\put(140,10.0){\line(0,1){90}}
\put(150,10.0){\line(0,1){90}}
\put(160,10.0){\line(0,1){90}}%
\put(175,30){\footnotesize $p+1$}
\end{picture}
\end{picture}
\ee
where
\be
\label{ni}
\tilde{s}_{i+1}\leq  n_i \leq  \tilde{s}_{i}\;
\ee
(with the convention that $ 2 \leq i \leq k$ and
$\tilde{s}_{k+1}=0$). In the $AdS_d$ case, Weyl tensors
(\ref{adweyl}) are components of some irreducible
$o(d-1,2)$-module. This implies that  Weyl tensors
(\ref{adweyl}) should be related to each other by some
differential equations which express compatibility conditions
(\textit{i.e.}, Bianchi identities) for the expressions of higher
spin curvatures in terms of Weyl tensors. Systematic analysis of
these relations to be presented elsewhere will lead to the full
unfolded formulation of the higher spin dynamics for free mixed
fields in $AdS_d$.

The Weyl tensors (\ref{adweyl}), (\ref{ni}) contain the primary
Weyl tensor
\be
\label{wtensor}
C^{a_0(s), \ldots , a_{p}(s), a_{p+2}(\tilde{s}_2),\ldots, a_{p+p_2}(\tilde{s}_2), \ldots, a_q(\tilde{s}_k) }\;
\ee
associated with the  Young tableau (\ref{adweyl})
having the minimal  possible number of cells

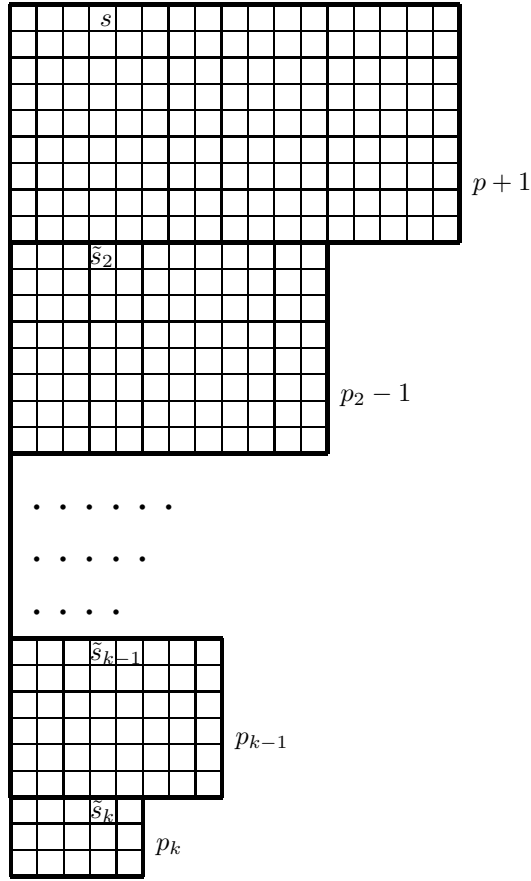
\be
\label{weyl}
\bigskip
\begin{picture}(185,350)%
\begin{picture}(109,300)%
{\linethickness{.500mm}
\put(30,32){\footnotesize $\tilde{s}_{k}$}
\put(50,10){\line(0,1){30}}
\put(00,10){\line(1,0){50}}
\put(00,10){\line(0,1){30}}
\put(00,40){\line(1,0){50}}}
\put(.0,40){\line(1,0){50}}
\put(.0,30){\line(1,0){50}}
\put(.0,20){\line(1,0){50}}
\put(10,10){\line(0,1){30}}
\put(20,10){\line(0,1){30}}
\put(30,10){\line(0,1){30}}
\put(40,10){\line(0,1){30}}
\put(55,20){\footnotesize $p_{k}$}
\end{picture}
\begin{picture}(125,90)(113,-30)%
{\linethickness{.500mm}
\put(30,62){\footnotesize $\tilde{s}_{k-1}$}
\put(80,10){\line(0,1){60}}
\put(00,10){\line(1,0){80}}
\put(00,10){\line(0,1){60}}
\put(00,70){\line(1,0){80}}}
\put(.0,60){\line(1,0){80}}
\put(.0,50){\line(1,0){80}}
\put(.0,40){\line(1,0){80}}
\put(.0,30){\line(1,0){80}}
\put(.0,20){\line(1,0){80}}
\put(10,10){\line(0,1){60}}
\put(20,10){\line(0,1){60}}
\put(30,10){\line(0,1){60}}
\put(40,10){\line(0,1){60}}
\put(50,10){\line(0,1){60}}
\put(60,10){\line(0,1){60}}
\put(70,10){\line(0,1){60}}
\put(85,30){\footnotesize $p_{k-1}$}
\end{picture}
\begin{picture}(75,75)(242,-90)%
{\linethickness{.5mm}
\put(00,10){\line(0,1){70}}}
\put(10,20){\circle*{2}}
\put(20,20){\circle*{2}}
\put(30,20){\circle*{2}}
\put(40,20){\circle*{2}}
\put(50,40){\circle*{2}}
\put(40,40){\circle*{2}}
\put(30,40){\circle*{2}}
\put(10,40){\circle*{2}}
\put(20,40){\circle*{2}}
\put(60,60){\circle*{2}}
\put(50,60){\circle*{2}}
\put(40,60){\circle*{2}}
\put(30,60){\circle*{2}}
\put(10,60){\circle*{2}}
\put(20,60){\circle*{2}}
\end{picture}
\begin{picture}(125,95)(321,-160)%
\put(30,82){\footnotesize $\tilde{s}_2$}%
{\linethickness{.500mm}
\put(120,10){\line(0,1){80}}
\put(00,90){\line(1,0){120}}
\put(00,10){\line(1,0){120}}
\put(00,10){\line(0,1){80}}}
\put(00,70){\line(1,0){120}}
\put(00,80){\line(1,0){120}}
\put(00,60){\line(1,0){120}}
\put(00,50){\line(1,0){120}}
\put(00,40){\line(1,0){120}}
\put(00,30){\line(1,0){120}}
\put(00,20){\line(1,0){120}}
\put(10,10.0){\line(0,1){80}}
\put(20,10.0){\line(0,1){80}}
\put(30,10.0){\line(0,1){80}}
\put(40,10.0){\line(0,1){80}}
\put(50,10.0){\line(0,1){80}}
\put(60,10.0){\line(0,1){80}}
\put(70,10.0){\line(0,1){80}}
\put(80,10.0){\line(0,1){80}}
\put(90,10.0){\line(0,1){80}}
\put(100,10.0){\line(0,1){80}}
\put(110,10.0){\line(0,1){80}}
\put(125,30){\footnotesize $p_2-1$}
\end{picture}
\begin{picture}(175,85)(450,-240)%
\put(34,92){\footnotesize $s$}
{\linethickness{.500mm}
\put(00,100){\line(1,0){170}}%
\put(00,10){\line(1,0){170}}%
\put(00,10){\line(0,1){90}}%
\put(170,10){\line(0,1){90}}}%
\put(00,90){\line(1,0){170}}
\put(00,80){\line(1,0){170}}
\put(00,70){\line(1,0){170}}
\put(00,60){\line(1,0){170}}
\put(00,50){\line(1,0){170}}
\put(00,40){\line(1,0){170}}
\put(00,30){\line(1,0){170}}
\put(00,20){\line(1,0){170}}
\put(10,10.0){\line(0,1){90}}
\put(20,10.0){\line(0,1){90}}
\put(30,10.0){\line(0,1){90}}
\put(40,10.0){\line(0,1){90}}
\put(50,10.0){\line(0,1){90}}
\put(60,10.0){\line(0,1){90}}
\put(70,10.0){\line(0,1){90}}
\put(80,10.0){\line(0,1){90}}
\put(90,10.0){\line(0,1){90}}
\put(100,10.0){\line(0,1){90}}
\put(110,10.0){\line(0,1){90}}
\put(120,10.0){\line(0,1){90}}
\put(130,10.0){\line(0,1){90}}
\put(140,10.0){\line(0,1){90}}
\put(150,10.0){\line(0,1){90}}
\put(160,10.0){\line(0,1){90}}%
\put(175,30){\footnotesize $p+1$}%
\end{picture}
\end{picture}
\ee
It can be interpreted as the invariant tensor of \cite{hull}
associated with the flat space mixed-symmetry field
described by the Young tableau with the minimal number of cells.
In the $AdS_d$ background all other Weyl tensors in (\ref{adweyl})
turn out to be  expressed through  derivatives of  the primary Weyl tensor
(\ref{weyl}) by virtue of Bianchi identities.

The primary Weyl tensor (\ref{wtensor})
parameterizes the components of the curvature
\be
\label{h2}
R_{(p+1)}{}^{a_0(s-1), \ldots , a_{p}(s-1), a_{p+2}(\tilde{s}_2),\ldots, a_{p+p_2}(\tilde{s}_2), \ldots, a_q(\tilde{s}_k)}(x)\;
\ee
defined by the formula
\be
\label{fow}
\ba{c}
C^{a_0(s), \ldots , a_{p}(s), a_{p+2}(\tilde{s}_2),\ldots, a_{p+p_2}(\tilde{s}_2), \ldots, a_q(\tilde{s}_k) }
(x)
\\
\\
= R^{[a_0 ...a_p];\,}{}^{a_0(s-1), \ldots , a_{p}(s-1), a_{p+2}(\tilde{s}_2),\ldots, a_{p+p_2}(\tilde{s}_2), \ldots, a_q(\tilde{s}_k)}(x)\,.
\ea
\ee

As follows from the analysis of  \cite{LV,Shayn,d5}, primary Weyl
tensors are classified by the cohomology group
$H^{p+1}(\sigma_- )$, where $\sigma_-$ is the part of the full
$AdS_d$ covariant derivative that decreases a number of Lorentz
indices   ($\sigma_-$ was called $\tau_-$ in \cite{LV}).
It has the following structure
\be
\label{tau-}
\sigma_- (R)^{a_0(r_0),\,\ldots \,,a_q(r_q)}=\sum_{i} \alpha_i (r_j)
{\cal
P}^{(i)}\Big(E_{a_i}\wedge R^{a_0(r_0),\,\ldots, a_i (r_i+1)\,,\ldots
\,,a_q(r_q)}\Big)\;,
\ee
where $\alpha_i (r_j)$ are some coefficients and
${\cal P}^{(i)}$ are  projectors that
guarantee that l.h.s. of (\ref{tau-}) is some  Young tableau
$Y(r_0,  \ldots, r_q)$ from the list
of Lorentz representations (\ref{lorblock}), (\ref{t})
associated with the system under consideration\footnote{For
 particular examples of mixed-symmetry fields
having Young symmetries of the types $Y(2,1)$ and $Y(2,2)$
the form of $\sigma_-$ can be easily read off from
 Eqs.(\ref{curv}), (\ref{wincurv}) of next section.}.
As a consequence of the flatness of the
$AdS_d$ covariant derivative (2.13), the operator $\sigma_-$
turns out to be nilpotent, $\sigma_-^2 =0$. (Note also that the operator
$\sigma_-$ is  Lorentz invariant.) Then one can see that
Bianchi identities require a part of the curvature associated with a
primary Weyl tensor to be $\sigma_-$
closed, {\it i.e.} to satisfy $\sigma_- (R)=0$. On the other hand any
$\sigma_-$ exact part of $R$ can be adjusted to zero by an appropriate
choice of constraints for auxiliary and extra fields.

It is elementary to see that the curvature (\ref{fow})
belongs to $H^{p+1}(\sigma_- )$ using a basis for Young tableaux
with antisymmetries associated with the columns manifest. Indeed,
the application of $\sigma_-$ (\ref{tau-}) may give a non-zero
result only if the index of the frame field is contracted with some of
the indices of the shortest columns of height $p+1$ (all other terms are
projected to zero by the projectors
in (\ref{tau-}) because any Young tableaux with a cell cut from any
other column is not in the list of the Lorentz representations
associated with the chosen field). In that case, the result is also
zero because of the Young property of Young tableaux in the
antisymmetric basis, which requires that antisymmetrization of all
indices of some column and any index from some other column of
less or equal
height gives zero. The antisymmetrization
over $p+2$ indices results from contraction with $p+2$ frame 1-forms.

Thus the curvature
(\ref{h2}), (\ref{fow})  is $\sigma_-$ closed. But it cannot be
$\sigma_-$ exact  because the Young tableau (\ref{weyl})
just does not appear among the Lorentz representations contained
in exact curvatures. Indeed, tensoring the form indices
with the tangent ones for any allowed $W$ in the exact representation
$R^{a_0(r_0),\,\ldots \,,a_q(r_q)}  =  \sigma_- (W)^{a_0(r_0),\,\ldots
  \,,a_q(r_q)}$ it is  possible to obtain tableaux with
at most $p$ cells in the last right column.

Note that  traces contained in the primary Weyl tensor
need separate consideration because traces of components
having different Young symmetry types may be related.
We therefore consider traceless
$C^{a_0(s), \ldots , a_{p}(s), a_{p+2}(\tilde{s}_2),\ldots, a_{p+p_2}(\tilde{s}_2), \ldots, a_q(\tilde{s}_k) }(x)
\in  B^{d-1,1}_{0}(\underbrace{s,\, ...\,,s}_{p+1},\underbrace{\tilde{s}_2,\,...\,,\tilde{s}_2}_{p_2-1},\ldots, \tilde{s}_k)$ only.

Thus, the frame-like   formulation of the mixed symmetry
massless higher spin fields we propose leads to the on-mass-shell
nontrivial primary Weyl tensor \\
$C^{a_0(s), \ldots , a_{p}(s), a_{p+2}(\tilde{s}_2),\ldots, a_{p+p_2}(\tilde{s}_2), \ldots, a_q(\tilde{s}_k)}(x)$
 invariant under $AdS_d$ higher spin gauge symmetries.
By construction, it contains $s-\tilde{s}_2$ derivatives of the physical
field\footnote{This is  because
additional indices in the curvature compared to the original
physical field are carried by the space-time
derivative operators which appear either through derivatives in the
higher spin curvature or  upon resolving constraints for auxiliary
and extra fields in terms of derivatives
of the physical fields.}
and is described by the Young tableau (\ref{weyl}).
Other tensors in (\ref{adweyl}) contain more (but no more than $s$)
derivatives of the physical field. They are expressed via derivatives
of the primary $AdS_d$ Weyl tensor but are expected to
become in the flat limit primary  Weyl tensors associated
with independent flat higher spin subsystems.

\section{Examples}

To illustrate the general scheme described in section 2, we first
consider two simplest examples of mixed-symmetry fields,
namely, a three-cell ``hook" field $\Phi^{a(2),\,b}(x)$ and a four-cell
``window" field $\Phi^{a(2),\,b(2)}(x)$. In each case we present
the full set of $p$-form gauge fields which consists of the
physical and auxiliary fields and build the actions which properly
describe irreducible $(A)dS_d$ dynamics\footnote{In the case of
$dS_d$ space-time we do not require the theory to describe unitary
dynamics with bounded energy.}. We derive the second-order equations of
motion on the
metric-type fields $\Phi^{a(2),\,b}(x)$ and
$\Phi^{a(2),\,b(2)}(x)$ which, in an appropriate gauge reproduce
the equations obtained by Metsaev \cite{Metsaev}.
The sets of fields we use in this section  are equivalent to those used by
Zinoviev in \cite{Zinoviev2} to describe ``hook" and ``window"
tableaux within first-order formalism. The universal $p$-form description
suggested in this paper makes higher spin symmetries manifest, however.
In the flat limit, the ``hook" theory yields an
additional symmetry not placed in the initial $(A)dS_d$
formulation. The case of ``window" tableau is generalized to
rectangular two-row Young tableau of an arbitrary length in
subsection 3.3. The action is uniquely fixed by {\it the extra field
decoupling condition}.

\subsection{Three-cell ``hook" tableau}

Consider Lorentz-covariant ``hook" field $\Phi^{[ab],\,c}(x)$
which is antisymmetric in the first two indices\footnote{In what follows,
we adopt for
``hook" and ``window" Young tableaux antisymmetric basis notations.
Namely, indices placed in square
brackets are assumed to be antisymmetrized as
$X^{[a}Y^{b]}=\frac{1}{2!}(X^aY^b-X^bY^a)$.}
$\Phi^{[ab],\,c}(x) = -\Phi^{[ba],\,c}(x)$  and satisfies the Young
symmetry condition
\be
\label{hook}
\Phi^{[ab,\,c]}(x) = 0\,.
\ee
The Lagrangian formulation for the metric-type field
$\Phi^{[ab],\,c}(x)$ in the flat background
was elaborated in \cite{curt,AKO}. The corresponding action is
invariant under the gauge transformation
\be
\label{hoogaugetr}
\delta\Phi^{[ab],\,c} = \d^a S^{bc}-\d^b S^{ac} + 2\d^c \Lambda^{[ab]}
-\d^a \Lambda^{[bc]}+\d^b \Lambda^{[ac]} \;
\ee
with antisymmetric gauge parameter $\Lambda^{[ab]}(x)=-\Lambda^{[ba]}(x)$
and symmetric gauge parameter $S^{ab}(x)=S^{ba}(x)$.
There is the level-2 gauge transformation with the gauge parameter
$\xi^a(x)$
\be
\label{2levelhook}
\delta S^{ab} =3(\d^a \xi^b + \d^b \xi^a)\;,\qquad
\delta \Lambda^{[ab]} = \d^b \xi^a - \d^a \xi^b\;.
\ee
The generalization of the flat theory to $AdS_d$ was constructed in
\cite{BMV}. It was shown that an appropriate deformation gives rise to
an action invariant under
\be
\label{3.3}
\delta\Phi^{[ab],\,c} = 2{\cal D}^c \Lambda^{[ab]}-{\cal D}^a
\Lambda^{[bc]}+{\cal D}^b \Lambda^{[ac]}\;,
\ee
while the gauge symmetry with the parameter $S^{ab}$
is lost in $AdS_d$. The absence of this gauge invariance
is in agreement with the fact that physical degrees of freedom of massless
$AdS_d$ fields are not described by irreps of Wigner little group
$o(d-2)$ \cite{BMV}.
For this particular example,  an  $AdS_d$ massless ``hook" field
decomposes in the flat
limit into a massless ``hook" field and a massless spin-2 symmetric
field.

To reformulate the $AdS_d$ theory of the
metric-type field $\Phi^{[ab],\,c}(x)$ \cite{BMV} within
 the scheme of  section 2 we introduce
the physical and auxiliary 1-forms
\be
\label{phaux}
e_{(1)}^{[ab]} = dx^{\underline{n}}\; e_{\underline{n}}{}^{[ab]}\;,
\qquad
\omega_{(1)}^{[abc]} = dx^{\underline{n}}\;
\omega_{\underline{n}}{}^{[abc]}
\ee
with antisymmetric tangent Lorentz indices.
Linearized curvature 2-forms associated with (\ref{phaux}) are
\be
\label{curv}
r^{[ab]}_{(2)} = {\cal D}e_{(1)}^{[ab]} + h_c\wedge\omega_{(1)}^{[abc]}\;,
\qquad
{\cal R}_{(2)}^{[abc]} = {\cal D}\omega_{(1)}^{[abc]} - 3\lambda^2\,
h^{[a}\wedge e_{(1)}^{bc]}\;,
\ee
where ${\cal D}$ is  a background Lorentz-covariant derivative. These curvatures
are invariant under the gauge transformations
\be
\label{gaugetr}
\delta e_{(1)}^{[ab]} = {\cal D}\xi_{(0)}^{[ab]} + h_c\xi_{(0)}^{[abc]}\;,
\qquad
\delta \omega_{(1)}^{[abc]} = {\cal D}\xi_{(0)}^{[abc]} - 3\lambda^2\,
h^{[a}\xi_{(0)}^{bc]}
\ee
with 0-form gauge parameters $\xi_{(0)}^{[ab]}$ and $\xi_{(0)}^{[abc]}$
antisymmetric in tangent indices. The gauge symmetry implies
the following Bianchi identities
\be
\label{Bianchi}
{\cal D}r_{(2)}^{[ab]} + h_c\wedge{\cal R}_{(2)}^{[abc]}=0\;,
\qquad
{\cal D}{\cal R}_{(2)}^{[abc]} - 3\lambda^2\, h^{[a}\wedge
r_{(2)}^{bc]}=0\;.
\ee
To see how the metric-type field $ \Phi^{[ab],c}(x)$ is encoded
in the gauge field $e_{\underline{n}}{}^{[ab]}(x)$ with the gauge law
(\ref{gaugetr}) we decompose
the Lorentz-covariant 1-form gauge fields (\ref{phaux}) into different
Young symmetry type components as
\be
\label{dec_e}
e_{\underline{n}}{}^{[ab]} \sim \Phi^{[ab],c}\oplus X^{[abc]}\;,
\ee
\be
\label{dec_omega}
\omega_{\underline{n}}{}^{[abc]} \sim \omega^{[abc],d}\oplus Y^{[abcd]}\;,
\ee
where the tensors $\Phi^{[ab],c}$ and $\omega^{[abc],d}$ contain
their traces.  The tensor $\Phi^{[ab],c} \in B_1^{d-1,1}(2,1)$
is identified with the dynamical metric-type field.
Its gauge transformation derived
from (\ref{gaugetr}) reads
\be
\label{phtr}
\delta \Phi^{[ab],c} = 2{\cal D}^c \xi_{(0)}^{[ab]}-{\cal D}^a
\xi_{(0)}^{[bc]}+{\cal D}^b \xi_{(0)}^{[ac]}
\ee
and is in agreement with (\ref{3.3}).
The totally antisymmetric component $X^{[abc]}$ of
$e_{\underline{n}}{}^{[ab]}(x)$
is compensated by the gauge shift generated by
the 0-form gauge parameter $\xi_{(0)}^{[abc]}$ (\ref{gaugetr}).

The Lorentz-covariant fields combine into a single 1-form field with $(A)dS_d$
tangent indices as
\be
\label{LorAds}
\Omega^{[ABC]}_{(1)} = e_{(1)}^{[ab]}\oplus \omega_{(1)}^{[abc]}\;,
\qquad
R^{[ABC]}_{(2)} =r^{[ab]}_{(2)}\oplus {\cal R}_{(2)}^{[abc]}\;,
\qquad
\xi^{[ABC]}_{(0)} = \xi^{[ab]}_{(0)}\oplus \xi_{(0)}^{[abc]}\;.
\ee
The gauge transformations, curvature and Bianchi identities
take now the form
\be
\delta\Omega_{(1)}^{[ABC]} = D_0\xi_{(0)}^{[ABC]}\;,
\qquad
R^{[ABC]}_{(2)}=D_0\Omega_{(1)}^{[ABC]}\;,
\qquad
D_0R_{(2)}^{[ABC]}=0\;,
\ee
where $D_0$ is the background $(A)dS_d$ derivative (\ref{zerocurv}),
(\ref{backderaction}).

The most general parity-invariant action is
written in terms of $(A)dS_d$ covariant tensors
\be
\label{acthook}
\ba{c} \dps
{\cal S}_2= \frac{\kappa_1}{\lambda^2}\int_{{\cal
M}^d}\;\epsilon_{ABCDM_5\ldots M_{d+1}}E^{M_5}\wedge\ldots \wedge
E^{M_d}\,V^{M_{d+1}}
\wedge R_{(2)}^{ABE}\wedge R_{(2)}^{CD}{}_{E}
\\
\\
\dps
+\frac{\kappa_2}{\lambda^2}\int_{{\cal
M}^d}\;\epsilon_{ABCDM_5\ldots M_{d+1}}E^{M_5}\wedge\ldots \wedge
E^{M_d}\,V^{M_{d+1}}
\wedge R_{(2)}^{ABE}\wedge R_{(2)}^{CDF}\,V_E\,V_F\;.
\\
\ea
\ee
Here $\kappa_{1,2}$ are arbitrary dimensionless constants.
By adding the total derivative term
\be
\ba{c}
\dps
\frac{1}{\lambda^2}\int_{{\cal
M}^d}\;d\Big(\epsilon_{ABCDEM_6\ldots M_{d+1}}E^{M_6}\wedge\ldots \wedge
E^{M_d}\,V^{M_{d+1}}
\wedge R_{(2)}^{ABC}\wedge R_{(2)}^{DE}{}_{F}\,V^F\Big)\;,
\ea
\ee
the freedom in $\kappa_{1,2}$ can be fixed up to
an overall multiplicative factor in front of the action (\ref{acthook}).
The variation of the action yields the following equations of motion:
\be
\label{eom_ads}
\frac{\delta{\cal S}_2}{\delta\Omega_{(1)}^{[ABC]}}=0\quad
\Leftrightarrow
\quad
\epsilon^{[AB}{}_{M_1 ... M_{d-1}}E^{M_3}\wedge\ldots\wedge E^{M_{d-1}}
\wedge R_{(2)}^{C]M_1M_2}\; = 0\;.
\ee
Converting all world indices into tangent ones, the equation of motion
(\ref{eom_ads}) can be rewritten
in terms of Lorentz-covariant components (\ref{LorAds}) as
\be
\label{dyn_eq}
\frac{\delta{\cal S}_2}{\delta e_{(1)}^{[ab]}}=0
\quad
\Rightarrow
\quad
({\cal R}_{st}{}^{[b\,;\,st}\delta^{a]}{}_{d}+2{\cal
R}_{dt}{}^{[b\,;\,ta]})\,=0\;,
\ee
\be
\label{constraint}
\frac{\delta{\cal S}_2}{\delta \omega_{(1)}^{[abc]}}=0
\quad
\Rightarrow
\quad
(\,r_k{}^{[c\,;\,bk} \, \delta^{a]}{}_{d}+r_d{}^{[c\,;\,ab]})=0\;.
\ee
The general solution for these linear restrictions on the
curvatures $r$ and ${\cal R}$
which conform Bianchi identities (\ref{Bianchi}) is
\be
\label{T}
r_{(2)}^{[ab]} = h_c\wedge h_d\; T^{[ab],[cd]}\;,
\ee
\be
\label{C}
{\cal R}_{(2)}^{[abc]} = h_d\wedge h_f \;C^{[abc],[df]}\;,
\ee
where  0-forms   $C^{[abc],\,[df]}$ and $T^{[ab],\,[cd]}$
are described by traceless two-column Young tableaux
\be
\label{trweyl}
C^{[abc],[df]}\,\eta_{cd}=0\;,
\qquad
\lambda^2\,T^{[ab],[cd]}\,\eta_{bc}=0\;.
\ee
They parameterize those components of the field strengths which can
be nonzero on-mass-shell and are called higher spin Weyl tensors.
The tensor $T^{[ab],[cd]}$ is the particular case
of  the primary Weyl tensor (\ref{weyl}). The tensor $C^{[abc],[df]}$,
corresponds to the additional Weyl tensors  (\ref{adweyl}) and can be
expressed as a first derivative of $T^{[ab],[cd]}$.
Note that in the limit $\lambda=0$
the tracelessness condition for $T^{[ab],[cd]}$ disappears.
This reflects the fact of appearance of an additional symmetry
in Minkowski space (see below).

The equation (\ref{constraint}) is the constraint which expresses
the auxiliary field in terms of derivatives of the physical
field. By gauge fixing $X^{[abc]}$ to zero (cf. (\ref{gaugetr}),
(\ref{dec_e})) one finds \be \label{aux} \omega^{[abc],d} =
-\frac{1}{2}\Big({\cal D}^a \Phi^{[bc],d}-{\cal D}^b
\Phi^{[ac],d}+ {\cal D}^c \Phi^{[ab],d}\;\Big). \ee For this
gauge it follows that $Y^{abcd} = 0 $ in (\ref{dec_omega}).
Substituting (\ref{aux}) into the equation (\ref{dyn_eq}) which
contains first derivatives of the auxiliary field, one finds the
second-order equation on the metric-type field $\Phi^{[ab],c}$ \be
\label{pheq} ({\cal D}^2  + ... + 3\lambda^2\,)\Phi^{[ab],c} =
0\;, \ee where terms containing ${\cal D}_c \Phi^{[ca],\,b},\,
{\cal D}_c \Phi^{[ab],\,c}$ or $\Phi^{[ab],\,}{}_{b}$ are omitted.
The covariant D'Alembertian is \be \label{dal} {\cal D}^2\equiv
{\cal D}_a{\cal D}^a=h^{\underline{n}}{}_a{\cal
D}_{\underline{n}}(h^{\underline{m}}{}^a{\cal
D}_{\underline{m}})\;, \ee where ${\cal D}_{\underline{m}}$ is
the background Lorentz-covariant derivative (\ref{Lorentzderiv}).
This equation coincides with that found in \cite{BMV} and, in the
covariant gauge $\Phi^{[ab],\,}{}_{b} = 0$, reproduces the
equation found by Metsaev \cite{Metsaev}.

By virtue of Bianchi identities (\ref{Bianchi}),
the second-order equation (\ref{pheq}) can be
equivalently rewritten in the form
\be
\label{trDF}
\eta_{ad}{\cal D}^{[a} F^{bc],df} = 0\;,
\ee
where $F^{[ab],[cd]} = {\cal D}^{[a} \Phi^{[cd],b]}+{\cal D}^{[c}
\Phi^{[ab],d]}$ is the $Y(2,2)$ projection
of the physical curvature $r_{(2)}^{[ab]}$ (\ref{curv}).
By construction, the tensor
$F^{[ab],[cd]}$ is invariant under the gauge transformations of the
metric-like field (\ref{phtr}). This implies that the second-order
equation of motion  (\ref{trDF}) which has the symmetries of the field
$\Phi^{[ab],c}$ is gauge invariant.

Now let us discuss the flat limit
of the action (\ref{acthook}) and equations (\ref{pheq}), (\ref{trDF}).
As is seen from the equations of motion (\ref{eom_ads}), all terms
containing
poles $1/\lambda$ enter the action through total derivatives.
As a result,
the action admits a well defined flat limit
at $\lambda\rightarrow 0$ (\ref{acthook}) upon adjusting
appropriate total derivative terms carrying negative
powers of $\lambda$.
Indeed, making use of the freedom in the parameters $\kappa_{1,2}$
(\ref{acthook}),
the action (\ref{acthook}) can be rewritten in the form
valid both for the flat and $(A)dS_d$ backgrounds
\be
\label{actLor}
\ba{c}
\dps
{\cal S}_2= \int_{{\cal M}^d}\;\epsilon_{abcdm_5\ldots m_{d}}
h^{m_5}\wedge\ldots \wedge h^{m_d} \wedge r_{(2)}^{ab}\wedge
r_{(2)}^{cd}\;.
\ea
\ee
For the flat space case one replaces ${\cal D}_{\underline{m}}
\rightarrow \d_{\underline{m}}$.
The equations of motion have the form   (\ref{T}) and
(\ref{C}) at $\lambda=0$.
However, a special feature of the theory in the flat limit
is that, in accordance with (\ref{hoogaugetr}),
an additional symmetry with the symmetric parameter $S^{ab}=S^{ba}$,
$S^a{}_a\neq 0$
appears in flat background \cite{BMV} for the second-order
field equations (\ref{pheq}). Note that
trivialization of the second equation in (\ref{trweyl}) in the
flat limit is just the Noether identity for this new symmetry.

It is worth to comment that in our formalism there is a systematic
way to show that the gauge invariance with the symmetric parameter
$S^{ab}$ appears in the flat limit by observing  that the
expression (\ref{aux}) for the auxiliary field
$\omega^{[abs],d}(\Phi)$ in terms of the metric-type field  turns
out to be invariant  under the gauge transformation with the
parameter $S^{ab}$ in the flat limit. The  existence of two types of
first-order invariant
expressions $F^{[ab],[cd]}(\Phi)$ and $\omega^{[abc],d}(\Phi)$ for
the ``hook" field which are invariant under gauge transformations
with antisymmetric and symmetric parameters, respectively, was
originally found in \cite{curt,Zinoviev}.
These tensors get natural geometric interpretation of particular
field strength and connection in our geometric approach. The difference in
their interpretation
is because only one of the two
types of symmetries remains unbroken in $AdS_d$.

\subsection{Four-cell ``window" tableau}

The mixed-symmetry  field described by the four-cell
rectangular ``window" tableau was  considered in
\cite{Pashnev,hull2,hull1} for  the flat background and in \cite{Zinoviev}
for
the $(A)dS_d$ background. Here we demonstrate how our approach
reproduces the analysis of \cite{Pashnev,Zinoviev}.

Consider Lorentz-covariant ``window" field $\Phi^{[ab],\,[cd]}(x)$
antisymmetric in the first and second groups of indices
and satisfying the Young symmetry condition
\be
\label{window}
\Phi^{[ab,\,c]d}(x) = 0\,.
\ee
The corresponding gauge symmetry is given by
\be
\label{gaugewind}
\delta\Phi^{[ab],\,[cd]} = {\cal D}^a S^{[cd],\,b}-{\cal D}^b
S^{[cd],\,a}+ {\cal D}^c S^{[ab],\,d}-{\cal D}^d S^{[ab],\,c} \;
\ee
with the parameter $S^{[ab],\,c}(x)$  antisymmetric in the first
two indices and satisfying the Young symmetry condition
$S^{[ab,\,c]}(x)=0$. The gauge parameter can be chosen either
traceless \cite{Pashnev} or not \cite{Zinoviev}. In the latter case
the gauge transformations (\ref{gaugewind}) are reducible: the
transformation $\delta S^{[ab],\,c} =2{\cal D}^c A^{ab} -{\cal D}^a
A^{bc}+{\cal D}^b A^{ac} $ with antisymmetric parameter $A^{[ab]}$
leaves invariant $\delta \Phi^{[ab],\,[cd]}$ (\ref{gaugewind}).

In accordance with the general prescription of section 2 introduce
the physical and auxiliary 2-form fields
\be
\label{winphaux}
e^{[ab]}_{(2)} = dx^{\underline{m}}\wedge dx^{\underline{n}}\;
e_{[\underline{m}\underline{n}]}{}^{[ab]}\;,
\qquad
\omega^{[abc]}_{(2)} = dx^{\underline{m}}\wedge dx^{\underline{n}}\;
\omega_{[\underline{m}\underline{n}]}{}^{[abc]}
\ee
with antisymmetric tangent Lorentz indices.
Linearized curvature 3-forms are
\be
\label{wincurv}
r^{[ab]}_{(3)} = {\cal D}e^{[ab]}_{(2)} + h_c\wedge\omega^{[abc]}_{(2)}\;,
\qquad
{\cal R}^{[abc]}_{(3)} = {\cal D}\omega_{(2)}^{[abc]} - 3\lambda^2\,
h^{[a}\wedge e^{bc]}_{(2)}\;.
\ee
They satisfy Bianchi identities
\be
\label{winBianchi}
{\cal D}r_{(3)}^{[ab]} + h_c\wedge{\cal R}_{(3)}^{[abc]}=0\;,
\qquad
{\cal D}{\cal R}_{(3)}^{[abc]} - 3\lambda^2\, h^{[a}\wedge
r_{(3)}^{bc]}=0\;
\ee
and are invariant under the gauge transformations
\be
\label{wingaugetr}
\delta e_{(2)}^{[ab]} = {\cal D}\xi_{(1)}^{[ab]} +
h_c\wedge\xi_{(1)}^{[abc]}\;,
\qquad
\delta \omega_{(2)}^{[abc]} = {\cal D}\xi_{(1)}^{[abc]} - 3\lambda^2\,
h^{[a}\wedge\xi_{(1)}^{bc]}\;
\ee
with the 1-form gauge parameters $\xi_{(1)}^{[ab]}$ and $\xi_{(1)}^{[abc]}$
antisymmetric in tangent indices. The gauge transformations
(\ref{wingaugetr})
are reducible. The corresponding transformations of the gauge parameters
$\xi_{(1)}^{[ab]}$ and $\xi_{(1)}^{[abc]}$ read
\be
\label{TRreduc}
\delta \xi_{(1)}^{[ab]} = {\cal D}\chi_{(0)}^{[ab]} +
h_c\chi_{(0)}^{[abc]}\;,
\qquad
\delta \xi_{(1)}^{[abc]} = {\cal D}\chi_{(0)}^{[abc]} - 3\lambda^2\,
h^{[a}\chi_{(0)}^{bc]}
\ee
with the level-2 $0$-form gauge parameters $\chi_{(0)}^{[ab]}$,
$\chi_{(0)}^{[abc]}$
antisymmetric in the tangent indices.

 Decompose  the 2-form gauge fields (\ref{winphaux}) into
\be
\label{windec_e}
e_{\underline{m}\underline{n}}{}^{[ab]} \sim \Phi^{[ab],[cd]}\oplus
X^{[abc],d}\oplus Y^{[abcd]}\;,
\ee
\be
\label{windec_omega}
\omega_{\underline{m}\underline{n}}{}^{[abc]} \sim
\omega^{[abc],[de]}\oplus Z^{[abcd],e}\oplus W^{[abcde]}\;,
\ee
where the tensors
$\Phi^{[ab],[cd]}$, $X^{[abc],d}$, $\omega^{[abc],[de]}$ and
$Z^{[abcd],e}$  have the Young symmetry types
$Y(2,2), Y(2,1,1), Y(2,2,1)$ and $Y(2,1,1,1)$, respectively
(and contain trace parts).
The components $X^{[abc],d}$ and $Y^{[abcd]}$  combine
into a single 1-form ${\cal E}_{(1)}^{[abc]}$
which  can be gauge fixed to zero by the  shift generated by
the 1-form gauge parameter $\xi_{(1)}^{[abc]}$ (\ref{wingaugetr}).
The remaining component  $\Phi^{[ab],[cd]}$ in (\ref{windec_e})
is identified with
the dynamical metric-type field and belongs to $B^{d-1,1}_{2}(2,2)$.
Its gauge transformation derived from (\ref{wingaugetr}) reads
\be
\label{winphtr}
\delta\Phi^{[ab],\,[cd]} = {\cal D}^a \xi^{[cd],\,b}-{\cal D}^b
\xi^{[cd],\,a}+ {\cal D}^c \xi^{[ab],\,d}
-{\cal D}^d \xi^{[ab],\,c} \;,
\ee
where the parameter $ \xi^{[ab],\,c}\in B^{d-1,1}_1(2,1)$, {\it i.e.}
is antisymmetric in the
first two indices, satisfies the Young symmetry condition $
\xi^{[ab,\,c]}=0$ and has non-vanishing trace  $ \xi^{[ab],}{}_b
\neq 0$. In fact, the parameter $ \xi^{[ab],\,c}$ is the component
of the 1-form gauge parameter $\xi_{(1)}^{[ab]}$ with its totally
antisymmetric part fixed to zero with the aid of the level-2 gauge
parameter $\chi_{(0)}^{[abc]}$ (\ref{TRreduc}).

The Lorentz-covariant component fields combine into the single 2-form field with
$(A)dS_d$
tangent indices as
\be
\ba{c}
\label{winLorAds}
\Omega^{[ABC]}_{(2)} = e^{[ab]}_{(2)}\oplus \omega_{(2)}^{[abc]}\;,
\quad
R^{[ABC]}_{(3)} =r^{[ab]}_{(3)}\oplus {\cal R}_{(3)}^{[abc]}\;,
\\
\\
\xi^{[ABC]}_{(1)} = \xi^{[ab]}_{(1)}\oplus \xi_{(1)}^{[abc]}\;,
\quad
\chi^{[ABC]}_{(0)} = \chi^{[ab]}_{(0)}\oplus \chi_{(0)}^{[abc]}\;.
\ea
\ee
The gauge transformations, curvature and  Bianchi identities
take the form
\be
\ba{c}
\delta\Omega_{(2)}^{[ABC]} = D_0\xi_{(1)}^{[ABC]}\;,
\quad
\delta\xi_{(1)}^{[ABC]} = D_0\chi_{(0)}^{[ABC]}\;,
\\
\\
R_{(3)}^{[ABC]}=D_0\Omega_{(2)}^{[ABC]}\;,
\quad
D_0R_{(3)}^{[ABC]}=0\;,
\ea
\ee
where $D_0$ is the background $AdS_d$ derivative
(\ref{zerocurv}), (\ref{backderaction}).

The parity-invariant gauge invariant action is uniquely fixed to the
form
\be
\label{wind}
\ba{c} \dps
{\cal S}_2= \frac{1}{\lambda^2}\int_{{\cal
M}^d}\;\epsilon_{ABCDEFM_7\ldots M_{d+1}}E^{M_7}\wedge\ldots \wedge
E^{M_d}\,V^{M_{d+1}}
\wedge R_{(3)}^{ABC}\wedge R_{(3)}^{DEF}\;.
\ea
\ee
Its variation gives rise to the equations of motion:
\be
\label{wineom_ads}
\frac{\delta{\cal S}_2}{\delta\Omega_{(2)}^{[ABC]}}=0\quad
\Leftrightarrow
\quad
\epsilon^{ABC}{}_{M_1 ... M_{d-2}}E^{M_4}\wedge\ldots\wedge E^{M_{d-2}}
\wedge R_{(3)}^{M_1M_2M_3} = 0\;.
\ee
Rewriting the equation (\ref{wineom_ads}) in Lorentz-covariant
 components one
finds
\be
\label{winC}
\frac{\delta{\cal S}_2}{\delta e_{(2)}^{[ab]}}=0
\quad
\Leftrightarrow
\quad
{\cal R}_{(3)}^{[abc]} = h_d\wedge h_e\wedge h_f \;C^{[abc],[def]}\;,
\ee
\be
\label{winT}
\frac{\delta{\cal S}_2}{\delta \omega_{(2)}^{[abc]}}=0
\quad
\Leftrightarrow
\quad
r_{(3)}^{[ab]} = 0\;,
\ee
where  the 0-form  $C^{[abc],\,[def]}$ is an arbitrary traceless tensor
\be
C^{[abc],[def]}\,\eta_{cd}=0\;
\ee
with the symmetry properties of the two-column Young tableau $Y(2,2,2)$.
This is the  Weyl tensor (\ref{weyl}).

The equation (\ref{winT}) is the constraint on the auxiliary
field which expresses it in terms of derivatives of the physical
field. Gauge fixing $X^{[abc],d}$ and $Y^{[abcd]}$ to zero
(\ref{windec_e}), one solves (\ref{winT}) as \be \label{winaux}
\omega^{[abc],[de]} = {\cal D}^a \Phi^{[bc],[de]}-{\cal D}^b
\Phi^{[ac],[de]}+ {\cal D}^c \Phi^{[ab],[de]}\;, \qquad
Z^{[abcd],e} = W^{[abcde]}=0\;. \ee Substituting this solution
into the equation (\ref{winC}) which contains first derivatives
of the auxiliary field, one finds the second-order equation on
the metric-type field $\Phi^{[ab],[cd]}$: \be \label{winpheq}
({\cal D}^2 + ... + (d+2)\lambda^2 )\Phi^{[ab],[cd]} =0\;, \ee
where ellipses denote terms containing ${\cal
D}_c\Phi^{[ca],[bd]}$ and $\Phi^a{}_{c\,,}{}^{[cb]}$. ${\cal D}^2$
is the covariant D'Alembertian given by (\ref{dal}). The equation
is invariant under the gauge transformations (\ref{winphtr}) and,
in the covariant gauge $\Phi^a{}_{c\,,}{}^{[cb]}=0$, reproduces
the equation found by Metsaev \cite{Metsaev}. As expected
\cite{BMV}, the flat limit of the field equation (\ref{winpheq})
yields no additional symmetries.

Note that the field $\omega^{[abc],[de]}(x)$
(\ref{windec_omega}) which appears as $Y(2,2,1)$ component of the
auxiliary 2-form connection in
our formalism, was introduced by Zinoviev in
\cite{Zinoviev2}. The Lagrangian of \cite{Zinoviev2} is
a particular gauge fixed version of (\ref{wind}).

\subsection{Two-row rectangular tableaux}

Consider now an arbitrary two-row rectangular Lorentz-covariant
bosonic
field propagating on
the flat or $(A)dS_d$ background. The consideration of the present
section is in many respects parallel to that
of  \cite{LV,d5} for totally symmetric higher spin fields.

Introduce 2-form gauge field which forms an  $(A)dS_d$
tangent tensor described by the
three row rectangular Young tableau $Y(s-1,s-1,s-1)$
\be
\label{adstworow}
\Omega^{A(s-1),B(s-1), C(s-1)}_{(2)} =dx^{\underline{m}}\wedge
dx^{\underline{n}}\;
\Omega_{[\underline{m}\underline{n}]}{}^{A(s-1),B(s-1), C(s-1)}\;,
\ee
subject to the tracelessness condition
\be
\label{trcondtworow}
\Omega^{A(s-1),B(s-1), C(s-1)}_{(2)}\;\eta_{A(2)}=0\;.
\ee
All other traces are also zero as a consequence of the
Young symmetry property.
In other words, $\Omega_{(2)}\in B^{d-1,2}_0(s-1,s-1,s-1)$.
The decomposition into the Lorentz-covariant higher spin tensors gives
\be
\label{3.45}
\Omega_{(2)}^{A(s-1),B(s-1), C(s-1)} \sim \;\; \bigoplus_{t=0}^{s-1}\;
\omega_{(2)}^{a(s-1),b(s-1), c(t)}\;,
\ee
where all Lorentz Young tableaux on r.h.s. of (\ref{3.45}) are
traceless. In accordance with the prescription of
 section 2, the Lorentz-covariant 2-form field with zero third row
($t=0$) is called physical, while those with non-zero third row
are auxiliary ($t=1$) or extra ($t>1$) and express via
derivatives of the physical field by virtue of certain
constraints. We do not discuss here the structure of
constraints for extra fields as they do not contribute to the
free equations of motion. The auxiliary field is expressed in
terms of the physical one by virtue of its field equation.

The gauge transformations are
\be\label{tworowgaugetr}
\delta\Omega_{(2)}^{A(s-1),B(s-1),C(s-1)} = D_0 \xi_{(1)}^{A(s-1),B(s-1),
C(s-1)}
\ee
with the 1-form gauge parameter $\xi_{(1)}$ subject to the same
irreducibility conditions as the field $\Omega_{(2)}$. In its turn, the
gauge
parameter $\xi_{(1)}$ has level-2 gauge transformation
\be
\delta \xi_{(1)}^{A(s-1),B(s-1), C(s-1)} = D_0 \chi_{(0)}^{A(s-1),B(s-1),
C(s-1)}
\ee
with the level-2 0-form gauge parameter $\chi_{(0)}$.

The linearized higher spin 3-form curvature
\be
\label{tworowcurv}
R_{(3)}^{A(s-1),B(s-1),C(s-1)} = D_0 \Omega_{(2)}^{A(s-1),B(s-1),C(s-1)}\;
\ee
is invariant under the gauge transformations (\ref{tworowgaugetr}).

In accordance with the general formula (\ref{fo})
the metric-type field is a part of the physical  field
\be
\label{mlfield2row}
\Phi^{a(s),b(s)} =\omega^{[ab];\,a(s-1),b(s-1)}\;,
\ee
where the 2-form world indices of the physical field are
converted into tangent ones. Other components of
$\omega_{(2)\;[\underline{m}\underline{n}]}{}^{a(s-1),\,b(s-1)}$
can be gauged away with the aid of the  shift gauge parameters
$\xi_{(1)\;\underline{m}}{}^{a(s-1),\,b(s-1),\,c}$. As a consequence of
the tracelessness condition (\ref{trcondtworow}) imposed on the
(\ref{adstworow}), the field
(\ref{mlfield2row}) satisfies the double-tracelessness conditions
\be
\label{doubletr}
\Phi^{a(s),b(s)}\eta_{a(2)}\eta_{a(2)}=0\;.
\ee
Thus $\Phi^{a(s),b(s)} \in B^{d-1,1}_2(s,s)$. From (\ref{doubletr})
it also follows that
\be
\label{doubletr2}
\Phi^{a(s),b(s)}\,\eta_{ab}\eta_{b(2)}=0\;,\qquad
\Phi^{a(s),b(s)}\,\eta_{a(2)}\eta_{b(2)}+2\,
\Phi^{a(s),b(s)}\,\eta_{ab}\eta_{ab}=0\;.
\ee
These are trace conditions of the work \cite{Pashnev},
where two-row mixed-symmetry fields on  Minkowski space were considered.

The gauge transformation law is
\be
\label{metrgaugetr}
\delta\Phi^{a(s),b(s)} = {\cal D}^b \Lambda^{a(s), b(s-1)}+(-)^s\,{\cal
D}^a \Lambda^{b(s),a(s-1)}\;.
\ee
Here the gauge parameter $\Lambda$ is defined as
\be
\label{metrgaugepar}
\Lambda^{a(s),b(s-1)} = \xi^{a;\,a(s-1),b(s-1)}\;,
\ee
and satisfies the trace conditions
\be
\label{metrgaugepar2}
\qquad  \Lambda^{a(s),b(s-1)}\,\eta_{a(2)}\eta_{a(2)} = 0\;,
\qquad \Lambda^{a(s),b(s-1)}\,\eta_{b(2)} = 0\;.
\ee
Being a consequence of the gauge law (\ref{tworowgaugetr}),
this definition is consistent with (\ref{doubletr}) and
(\ref{metrgaugetr}).
In accordance with general consideration of section 2
we see that $\Lambda^{a(s),\,b(s-1)} \in B^{d-1,1}_1(s,s-1)$.

The metric-type gauge field (\ref{mlfield2row})  subject
to the trace conditions (\ref{doubletr}) is analogous
to that considered in \cite{Pashnev}.
The field trace conditions of \cite{Pashnev}
arise automatically in our approach as a consequence of the irreducibility
of the
2-form in the tangent indices.
The difference however
is that the gauge parameter is not required to be traceless.
Instead,  weaker trace conditions (\ref{metrgaugepar2}) are imposed,
{\it i.e.} we have more gauge symmetries manifest in our approach.

Let us look for a parity-invariant action in the form
\be
\label{tworowaction}
\ba{c} \dps
{\cal S}_2= \half \int_{{\cal
M}^d}\;\sum_{p=0}^{s-2} a(s,p)\,\epsilon_{A_1 \ldots
A_{d+1}}E^{A_7}\wedge\ldots \wedge
E^{A_d}\,V^{A_{d+1}} V_{D_1}\cdots V_{D_{2(s-p-2)}}
\\
\\
\wedge R_{(3)}^{A_1B(s-2),A_2C(s-2),A_3D(s-2-p)F(p)}\wedge
R_{(3)}{}^{A_4}{}_{B(s-2),}{}^{A_5}{}_{C(s-2),}{}^{A_6D(s-2-p)}{}_{F(p)}\;,
\ea
\ee
where arbitrary coefficients $a(s,p)$ should be fixed by the extra field
decoupling condition. This action makes sense for $d\geq 6$.
Its general variation is
\be
\label{tworowvar}
\ba{c}
\dps
\delta{\cal S}_2 =  \frac{(-)^d\lambda}{(d-5)}\int_{{\cal
M}^d}\;\sum_{p=0}^{s-2} \Big( \frac{(s-p+1)(d-9+2(s-p))}{(s-p-1)}a(s,p)+
(s-p-1)a(s,p-1) \Big)
\\
\\
\dps
\times
\epsilon_{A_1 \ldots A_{d+1}}E^{A_6}\wedge\ldots \wedge
E^{A_d}\,V^{A_{d+1}} V_{D_1}\ldots V_{D_{2(s-p)-3}}
\\
\\
\dps
\wedge\Big( R_{(3)}^{A_1B(s-2), A_2C(s-2), D(s-p-1)F(p)}\wedge
\delta
\Omega_{(2)}^{A_3}{}_{B(s-2),}{}^{A_4}{}_{C(s-2),}{}^{A_5D(s-p-2)}{}_{F(p)}
\\
\\
- \, \delta\Omega_{(2)}^{A_1B(s-2), A_2C(s-2), D(s-p-1)F(p)}\wedge

R_{(3)}^{A_3}{}_{B(s-2),}{}^{A_4}{}_{C(s-2),}{}^{A_5D(s-p-2)}{}_{F(p)}\Big)\;.
\ea
\ee
To impose the extra field decoupling condition one should require
all terms in (\ref{tworowvar}) to vanish except for that with $p=0$. This
requirement
fixes the coefficients $a(s,p)$ up to a normalization factor
$\tilde{a}(s)$:
\be
a(p,s) =
\tilde{a}(s)\,(-)^p\frac{(s-p)(s-p-1)(d-11+2(s-p))!!}{(s-p-2)!}\;.
\ee
With these coefficients $a(p,s)$
the action depends essentially only on
\be
\Omega^{A(s-1),B(s-1)}_{(2)} \equiv
\Omega^{A(s-1),B(s-1),C(s-1)}_{(2)}\;V_{C(s-1)},
\ee
which is automatically $V$-transversal, and on the $V$-transversal part of
\be
\Omega^{A(s-1),B(s-1),C}_{(2)} \equiv
\Omega^{A(s-1),B(s-1),CD(s-2)}_{(2)}\;V_{D(s-2)}\;.
\ee
These are, respectively, the physical ($t=0$) and the auxiliary ($t=1$)
fields. The extra fields do not contribute into the free action
as guaranteed by the extra field decoupling condition. The auxiliary
field $\omega_{(2)}^{a(s-1),b(s-1),c}$ is expressed in terms of the first
derivatives of the physical field $\omega_{(2)}^{a(s-1),b(s-1)}$ by virtue
of its equation of motion.
Insertion of the expression for the auxiliary field
$\omega_{(2)}^{a(s-1), b(s-1),c}$ back into the action gives rise
to the higher spin action expressed entirely in terms of the
metric-type field (\ref{mlfield2row}) and its first derivatives. The gauge
invariance
is inbuilt by construction.
The flat limit does not yield any additional gauge symmetries.
This is in agreement with the general analysis of \cite{BMV},
where the class of Poincare irreps described by rectangular tableaux
was argued to admit an $AdS_d$ deformation.
Therefore, the resulting action possesses
correct higher spin gauge symmetries and describes properly
both  Minkowski and $(A)dS_d$ free dynamics.

\section{Conclusion}

The general approach proposed in this paper provides
manifestly gauge invariant framework for the formulation of
the dynamics of mixed-symmetry massless higher spin gauge
fields in (anti) de Sitter and flat space. As demonstrated
by the particular examples, the realization of the relevant
sets of higher spin fields in terms of $p$-forms taking
values in certain irreducible representations of the
$(A)dS$ algebras simplifies analysis considerably and looks
promising for the description of the Lagrangian dynamics of
a general mixed-symmetry field in $(A)dS_d$. Because higher
spin gauge forms introduced in this paper should result
from gauging of some non-Abelian  higher spin
symmetries the proposed approach  provides an important
information on the structure of  underlying higher spin
algebras.

Let us note that our approach gives less components for a generic
mixed-symmetry metric-type field compared to other examples
considered in the literature \cite{Labastida} within local
formulation of higher spin dynamics. This phenomenon takes place
starting from the first nontrivial example of the
field of the symmetry type $Y(3,2)$. It remains to see whether
this indicates some type of
reducibility of models of Ref. \cite{Labastida}, or is
a matter of particular gauge
fixing, or is specific for higher spin dynamics in $AdS_d$.
The same time we have more
higher spin gauge symmetries manifest compared to some
other examples \cite{Pashnev}.

\vspace{1.5cm}

\newpage
{\bf Acknowledgements}
\\
\\

M.V. is grateful to Laurent Baulieu  for warm hospitality at L.P.T.H.E.,
Universites Pierre et Marie Curie,
where a part of this work was done.
This work is supported by grants RFBR No 02-02-17067,
LSS No 1578-2003-2, INTAS No 00-01-254. The work of A.K. is partially
supported by grants MAC No 03-02-06462 and the Landau Scholarship
Foundation,
Forschungszentrum J\"u\-lich. The work of O.Sh. is partially
supported by grants MAC No 03-02-06465 and the Landau Scholarship
Foundation,
Forschungszentrum J\"u\-lich.

\end{document}